\documentclass[fleqn,10pt]{wlscirep}
\usepackage[utf8]{inputenc}
\usepackage[T1]{fontenc}
\usepackage{tabularx}
\usepackage{soul}
\usepackage{booktabs, makecell}
\usepackage{graphicx}
\usepackage{xcolor}
\usepackage{xspace}
\usepackage{colortbl}
\usepackage{subcaption}

\title{Dream Content Discovery from Reddit with an Unsupervised Mixed-Method Approach}

\author[1]{Anubhab Das}
\author[2,*]{Sanja \v{S}\'{c}epanovi\'{c}}
\author[3,4]{Luca Maria Aiello}
\author[5]{Remington Mallett}
\author[6]{Deirdre Barrett}
\author[2]{Daniele Quercia}
\affil[1]{Heritage Institute of Technology, India}
\affil[2]{Nokia Bell Labs, United Kingdom}
\affil[3]{IT University of Copenhagen, Denmark}
\affil[4]{Pioneer Centre for AI, Denmark}
\affil[5]{Northwestern University, USA}
\affil[6]{Harvard University, USA}
\affil[*]{Corresponding author (\texttt{sanja.scepanovic@nokia-bell-labs.com})}


\begin{abstract}
Dreaming is a fundamental but not fully understood part of human experience that can shed light on our thought patterns. Traditional dream analysis practices, while popular and aided by over 130 unique scales and rating systems, have limitations. Mostly based on retrospective surveys or lab studies, they struggle to be applied on a large scale or to show the importance and connections between different dream themes. To overcome these issues, we developed a new, data-driven mixed-method approach for identifying topics in free-form dream reports through natural language processing. We tested this method on 44,213 dream reports from Reddit's r/Dreams subreddit, where we found 217 topics, grouped into 22 larger themes: the most extensive collection of dream topics to date. We validated our topics by comparing it to the widely-used Hall and van de Castle scale. Going beyond traditional scales, our method can find unique patterns in different dream types (like nightmares or recurring dreams), understand topic importance and connections, and observe changes in collective dream experiences over time and around major events, like the COVID-19 pandemic and the recent Russo-Ukrainian war. We envision that the applications of our method will provide valuable insights into the intricate nature of dreaming.
\end{abstract}
\begin{document}

\newcommand{\rdreams}{\texttt{r/Dreams}\xspace\space}

\flushbottom
\maketitle

\section*{Introduction}

Dreaming is a fundamental human experience and a cornerstone of sleep psychology, yet its underlying mechanisms remain elusive. The fascination with the contents and meaning of our dreams dates back to early human civilizations,\cite{mota2020dream} but despite significant progress in dream research, fundamental questions about the physiological and psychological functions of dreaming remain unanswered, leaving us to ponder the question: \emph{why do we dream?}\cite{schredl2010characteristics} Previous literature on dream analysis suggests that dreaming plays a vital role in learning processes,\cite{hobson1977brain} has psychotherapeutic effects,\cite{hartmann1995making} and safeguards the brain's neuroplasticity.\cite{eagleman2021defensive} Recent AI-inspired theories have drawn parallels between the brain's activity during sleep and the functioning of artificial neural networks. One theory suggests that dreams act as ``garbage collectors'' that clear the memory,\cite{crick1983function} while another posits that they prevent over-fitting of the brain's neuronal network.\cite{hoel2021overfitted}

While the question of why we dream remains open, before we can even begin to answer it, we must first understand the nature of \emph{what we dream}. This question is important not only for helping us understand the fundamental function of dreams but also as it offers a window into our psyche and what is prominent in people's minds. Dream content is composed of fragments of waking life experiences and events,\cite{schredl2010dream} but these are not veridical replays, \cite{payne2010memory,wamsley2010dreaming} nor do they represent the entirety of dream content.\cite{revonsuo1995content} One popular approach to investigating dream content systematically is through \emph{content analysis}.\cite{schredl2010dream} This is a family of methods that analyze quantitatively the elements present in dreams to answer specific questions, such as whether depressed individuals experience more rejection in their dreams,\cite{beck1961dreams} how dream content changes from teenage years to adulthood,\cite{fogli2020our} or whether in the times of collective health crises there is a shift in the medical symptoms people dream about.\cite{vscepanovic2022epidemic} Although these studies may seem narrow in scope, they provide a critical foundation for addressing the overarching question of why we dream. The importance of content analysis is evidenced by the development of over 130 scales and rating systems for dream content analysis.\cite{winget1979dimensions} Early scales tended to score based on the raters' subjective interpretation of dream symbolism and rarely reported inter-rater reliability.\cite{hall1966content} Dream research became more systematic with the development of the \emph{Hall and van de Castle method} \cite{hall1966content} of dream content analysis, a quantitative system that scores dream reports based on the frequency and type of characters, interactions, activities, emotions, settings, and objects present in the dream. This method relies solely on the dream reports and does not use any additional information provided by the dreamer. Studies using the Hall and van de Castle scales have revealed consistent patterns and norms in dream content across different groups of people, such as gender, age, culture, and personality. For instance, women tend to dream more about family members, friends, and indoor settings, while men tend to dream more about strangers, violence, and outdoor settings. Moreover, research using these scales has demonstrated that dream content correlates with an individual's waking concerns and interests, such as work, relationships, hobbies, and fears.\cite{fogli2020our}

\paragraph{Limited content scope and representatives of existing scales.} Traditionally, dream researchers had to manually sift through large numbers of dream reports to gain insights into the range of dream topics.\cite{hall1966content} Even with recent developments in automated dream analysis,\cite{elce2021language} most studies continued to rely on experimenter-driven content searches, i.e., \emph{supervised approaches} for content analysis, such as the Hall and van de Castle method. These methods involve the use of \emph{predetermined} categories that are often biased towards existing knowledge of dreams. For example, dreams are often characterized as bizarre, involving impossible or improbable events that deviate from everyday experiences,\cite{colace2003dream} which may not fit within the predetermined categories established in the literature. As a result, current approaches to dream analysis may miss important aspects of dream content that fall outside of these predetermined categories. In contrast to traditional methods of dream coding, \emph{unsupervised theme discovery} from dream reports may provide a fresh perspective and a more comprehensive understanding of the categorization of dream content and its relationship with waking life events. 
Furthermore, previous studies relied on dream reports collected through \emph{retrospective surveys} that are susceptible to memory biases, and \emph{laboratory studies} that may be confounded by the strong impact of the laboratory setting on dream content.\cite{picard2021dreaming} Therefore, a more \emph{ecologically valid approach} to studying dream content at the population level is necessary.

\paragraph{Our unsupervised mixed-method approach for dream content discovery.} Our study addresses the identified research gap (i.e.,  limited scope and representatives of dream content scales) by developing an unsupervised mixed-method approach for dream content discovery, and by applying it to new large-scale data that more closely approximates spontaneous dream recollections than survey studies. To achieve this, we i) leveraged recent advances in AI for Natural Language Processing (NLP) and ii) used a crowd-sourced dataset of dream self-reports from the \rdreams community on Reddit. Unlike traditional lab studies, dream experiences shared on \rdreams are reported voluntarily and spontaneously, enabling us to collect a large set of dream reports and conduct an ecological study. We collected over 44K dream reports from more than 34K Reddit users over the past five years, and applied the BERTopic content analysis method \cite{devlin2018bert} to automatically discover topics in each dream report. The resulting taxonomy includes 22 \emph{themes} that can be broken down into 217 more specific \emph{topics}. Confirming its validity, we found that most of the themes in our taxonomy align with the dream element categories present in the Hall and Van de Castle scale, but the specific topics inside those themes provide a description of dreams that is much more detailed. Going beyond what was possible with existing scales, our method also allowed to uncover the importance and relationships among specific topics and themes. 

To demonstrate the applications enabled by our method, we used the metadata from the \rdreams community, and classified dreams into four types: \emph{nightmares, recurring dreams, vivid dreams}, and \emph{lucid dreams}. Our analysis revealed that each type of dream has distinct characteristics and prominent topics. Notably, nightmares were associated with \emph{scary} and \emph{shadowy} imagery, and \emph{sexual-assault} scenes. Vivid dreams featured rich expressions of \emph{feelings} and topics inducing extreme emotions such as \emph{pregnancy and birth}, \emph{religious} figures, \emph{war}, and \emph{aliens}. Lucid dreams were characterized by topics of \emph{control}, and by an overarching theme of \emph{mental reflections and interactions}. For recurring dreams, we found that the most salient topics were \emph{dating}, \emph{sex}, and \emph{cheating}, with recurring themes related to \emph{school} and mentions of parts of the \emph{human body}. Additionally, we investigated the relationship between dream topics and real-life experiences by studying the evolution of topics over the past five years. Our findings showed that the COVID-19 outbreak coincided with a gradual and collective shift in dream content. People started to dream less about \emph{people and relationships}, \emph{feelings}, \emph{sight and vision}, \emph{outdoor locations}, and \emph{movement and action}, and more about \emph{the human body, especially teeth and blood}, \emph{violence and death}, \emph{religious and spiritual} themes, and \emph{indoor locations}. Similarly, after the war in Ukraine started the topics about \emph{soldiers} and \emph{nuclear war} both peaked.

\section*{Results}\label{sec:results}

\begin{figure}[!ht]
\centering
\includegraphics[width=0.98\textwidth]{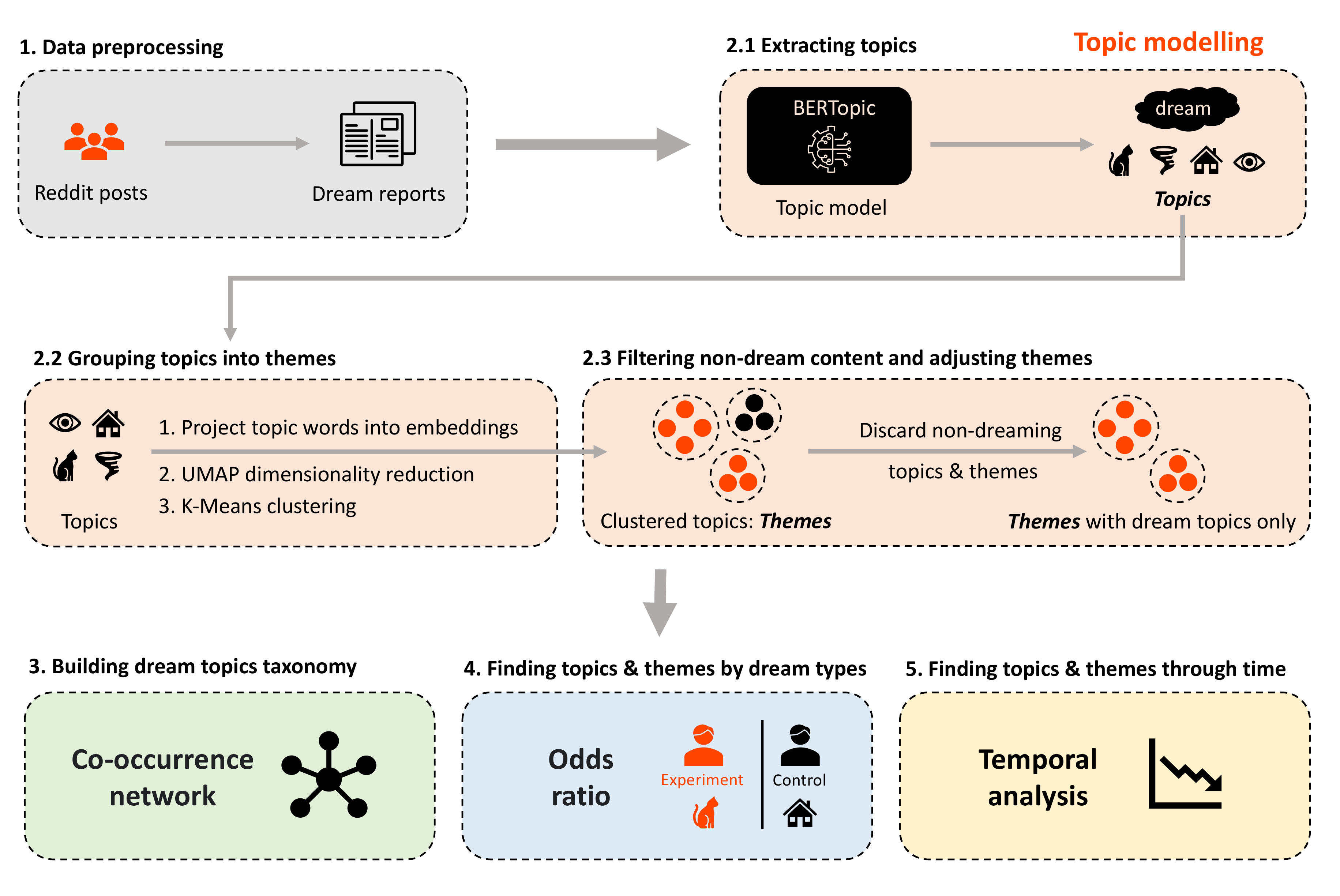}
\caption{\textbf{The framework of our study.} \emph{Stage 1.} involves data creation, while \emph{Stage 2.} steps are the core of our proposed unsupervised mixed-method approach for dream content discovery. \emph{Stages 3.-5.} demonstrate various applications enabled by our methodology, such as \emph{3.} uncovering the topic taxonomy in a particular dream report collection, \emph{4.} finding topics and themes that are specific to a given subset of dream reports in the given collection, and \emph{5.} tracking topic and theme trends through time.}\label{fig:methodology}
\end{figure}

Figure \ref{fig:methodology} outlines the framework of our study. It consist of three main stages: 1) \emph{data preprocessing}, which includes collecting dream reports, ideally in an ecological setting, and using NLP methods for cleaning the content; 2) \emph{topic modelling}, which is the core NLP stage for topic modelling. Once the topics and themes in the dream reports collection are discovered, various applications are supported, and we demonstrate three of those: 3) \emph{building a dream topics taxonomy}, which allows to uncover the relationships between individual themes and topics, as well as the frequency of each of them in the dream collection; 4) \emph{finding topics and themes by dream types}, which as an application that uses a proposed measure of topic or theme importance in a dream and odds-ratio analysis to discover topics that are specific to a dream type (or any other dream reports subset of choosing), and 5) \emph{finding topics and themes through time}, which is an application using the proposed topic or theme importance in a dream to quantify the prevalence of dream topics and themes through time.

\subsection*{Dream reports from Reddit}\label{sec:data}

In the established literature of dream analysis, dream reports are defined as \emph{``the recollection of mental activity which has occurred during sleep.''}\cite{schredl2010characteristics} To gather a dataset of such reports, we turned to Reddit, a social media platform organized in communities known as \emph{subreddits}. Using the PushShift API\cite{baumgartner2020pushshift}, we collected data from \rdreams, a subreddit where members share their dreams and engage informally in their interpretation --- which is common in therapy contexts~\cite{keller1995use} --- as well as in providing social and emotional support to other members.

\begin{figure*}
\centering
\begin{subfigure}{\textwidth}
  \centering 
    \includegraphics[width=0.99\textwidth]{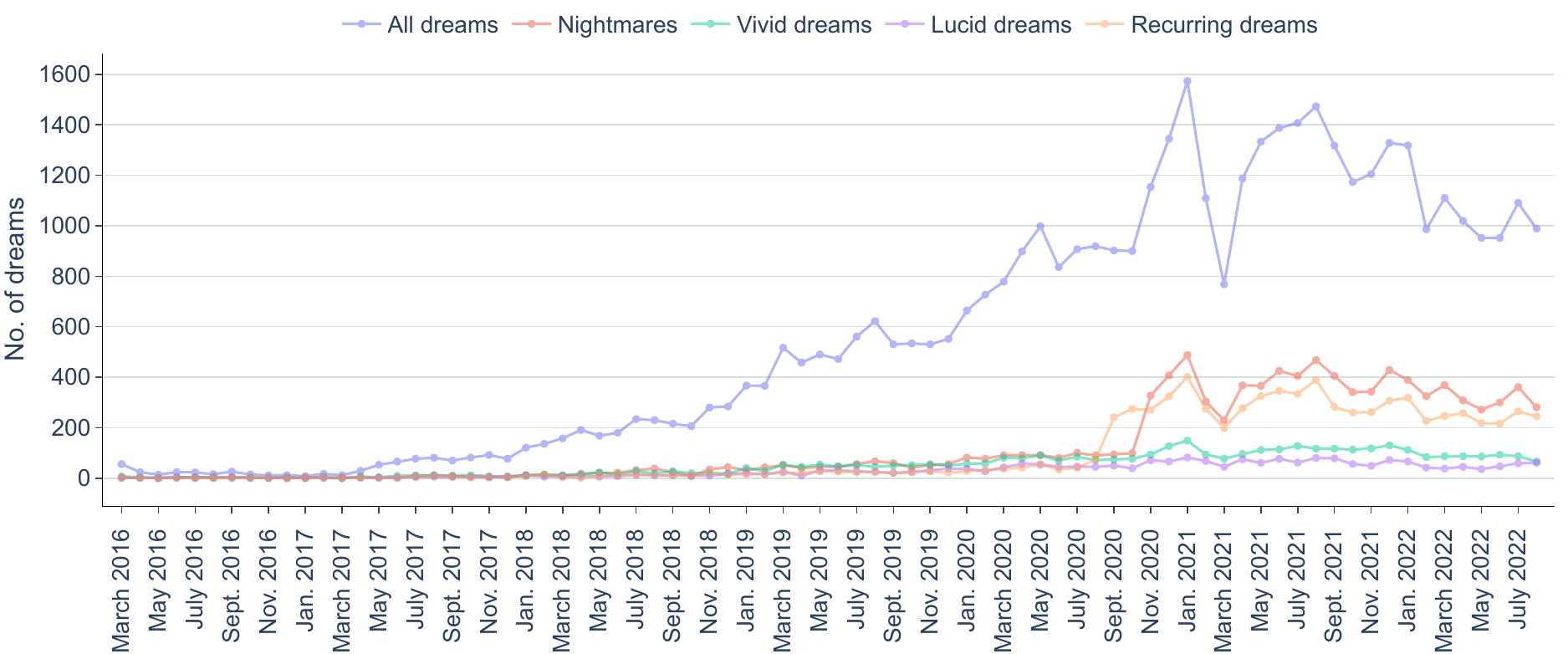}
    \caption{Monthly volume of the dream reports we collected from \rdreams, together with the volume of dream reports labeled with the four dream types used by the community.\label{fig:stream}}
\end{subfigure}%

\begin{subfigure}{\textwidth}
  \centering 
    \includegraphics[width=0.99\textwidth]{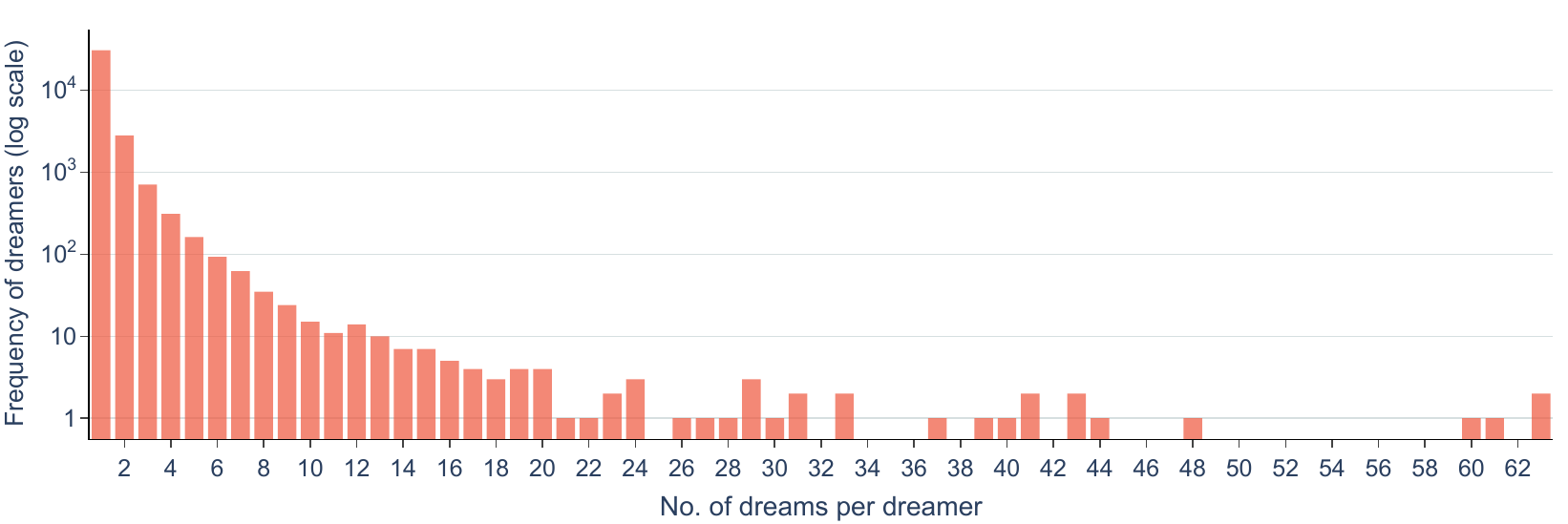}
    \caption{Distribution of number of dreams per user. The majority of users posted less than 10 dreams during the 5 years for which we collected the data. There is only a dozen of prolific dreamers who reported more than 30 dreams.}
    \label{fig:dreams_per_user}
\end{subfigure}
\caption{Dream reports statistics: (a) temporal, and (b) per dreamer.}
\end{figure*}

The \rdreams subreddit was established in September 2008, and as of June 2022, it had accumulated 280K subscribers. We collected over 185K posts published on \rdreams from March 2016 to September 2022; prior to 2016, the community was almost inactive. Authors on \rdreams annotate their posts with one or more tags selected from a fixed set community-specific labels called \emph{flairs}. Flairs on \rdreams denote posts that contain dream reports of a given type  (\emph{Short Dream}, \emph{Medium Dream}, \emph{Long Dream}, \emph{Nightmare}, \emph{Recurring Dream}, and \emph{Lucid Dream}) or posts that contain discussions about dreams in general (e.g., \emph{Dream Help}, \emph{Dream Art}, or \emph{Question}). To ensure that we considered only posts that contained dream reports, we only kept the 44,213 unique posts tagged with dream-type flairs (Figure~\ref{fig:stream}). In our analysis, a  \emph{dream report} was the concatenation of the title and body of each of these posts.

\begin{table}[!ht]
\centering
\small
\caption{Dream reports statistics; time period for which we collected the data is from March 2016 to September 2022.\label{tab:stats}}
\begin{tabular}{@{}p{.17\textwidth}p{.10\textwidth}p{.10\textwidth}p{.10\textwidth}p{.10\textwidth}p{.10\textwidth}@{}}
\toprule
\textbf{Dream type} & \textbf{All dreams} & \textbf{Vivid dreams} & \textbf{Nightmares} & \textbf{Lucid dreams} & \textbf{Recurring dreams} \\ \midrule
\# dream reports & 44213 & 3962 & 9823 & 2272 & 7597 \\ \midrule
\# users & 35006 & 3819 & 8959 & 2107 & 7201 \\ \midrule
Mean \# sentences & 17 $\pm$ 18 & 25 $\pm$ 23 & 20 $\pm$ 20 & 27 $\pm$ 27 & 14 $\pm$ 16 \\ \midrule
Mean \# words & 290 $\pm$ 285 & 434 $\pm$ 382 & 341 $\pm$ 323 & 465 $\pm$ 424 & 253 $\pm$ 256 \\ \midrule
Mean \# characters & 1455 $\pm$ 1444 & 2192 $\pm$ 1951 & 1713 $\pm$ 1640 & 2343 $\pm$ 2152 & 1274 $\pm$ 1299 \\ \midrule
Median  \# sentences  & 12 & 18 & 14 & 19 & 10 \\ \bottomrule
Median \# words & 205 & 325 & 252 & 338 & 178 \\ \midrule
Median \# characters & 1023 & 1636 & 1258 & 1697 & 895 \\ \midrule
\end{tabular}
\end{table}

\subsection*{Our unsupervised mixed-method approach for dream content discovery} 

As mentioned above, the core part of our approach shown in Figure \ref{fig:methodology} is the unsupervised mixed-method \emph{topic modelling} and it involves: 2.1) \emph{extracting topics} using an advanced NLP topic model, such as BERTopic \cite{grootendorst2022bertopic}; 2.2) \emph{grouping topics into themes} using a clustering method to group embedding representations of individual topics discovered in the previous step into themes, and 2.3) \emph{filtering non-dream content and adjusting themes}, which is a mixed-method part of the proposed methodology, requiring human, ideally dream experts knowledge to filter out topics or themes that do not pertain to the actual dream content, as well as to adjust any topic to theme associations, if needed. For more details about any of the steps of our approach, please refer to Methods section.

\subsection*{The Reddit taxonomy of dream topics}\label{sec:topics}

Using state-of-the-art topic modeling techniques, we have identified 217 semantically-cohesive topics that emerged from the dream reports analyzed (details in Methods). We assigned to each dream report the list of unique topics that our model was able to extract from the report text. The distribution of number of dreams associated with a topic is broad (Figure SI5), with only $38$ topics being associated with at least $1000$ dream reports. Table~\ref{tab:topics} shows the $20$ most frequent topics.

To offer a more concise representation of dream topics, we automatically clustered the 217 fine-grained topics into 22 \emph{themes} (see Methods), which we then manually parsed to assess their conformity to categories from the dream coding system by Hall and Van de Castle (\emph{HVdC} for brevity). \emph{HVdC} defines 12 categories and several subcategories of elements empirically relevant for quantitative dream analysis. All themes but one (a miscellaneous theme containing diverse topics) matched some \emph{HVdC} category or subcategory (Table \ref{tab:topic_clusters}).

At a high level of abstraction, \emph{HVdC} views the dream as a cast of characters (i), interacting with each other (ii), while being immersed in some background setting (iii). These three aspects emerged in the most prominent themes extracted from Reddit dream reports. The largest theme, both in terms of the number of topics it contains (17) and the number of dream reports associated with it (20K), is \emph{People and relationships}. The main topics included in this theme are \emph{family} members and relationships, and \emph{intimacy and romance} (see Table \ref{tab:topic_clusters} and Table 1 SI). Characters and interactions are also represented in themes concerning \emph{Animals}, \emph{Supernatural entities}, and \emph{Religious and Spiritual}, which map directly to the two HVdC subcategories of Animals and Imaginary Characters. Interactions respectively of aggressive and friendly type are represented in the themes \emph{Violence and Death} and \emph{Life Events}. The second most frequent group of themes involves events or elements that are typical of well-characterized places, such as home or the workplace. Among them, the most prominent theme is \emph{Indoor locations}, which includes 16 topics such as \emph{house}, \emph{hospital}, and \emph{mall}. These themes map to different subcategories of the \emph{HVdC} category of Settings.

The remaining themes correspond to the categories of Activities, Objects, Emotions, and Time in the \emph{HVdC} framework. With respect to Activities, Reddit users more frequently reported their mental activities and perceptions, rather than physical actions, when recounting their dreams. Specifically, visual and auditory sensory experiences were often recounted. The most commonly described categories of Objects included body parts (often with gruesome details) and personal objects, with a particular emphasis on technological devices such as phones or elements from virtual worlds of computer games. Emotions were represented by a single theme, characterized by common formulas for describing emotions of any type, with negative emotions being more prominently featured. Finally, while one theme captured temporal scales, such references were infrequent, appearing in fewer than 1,000 dreams.

Some of the \emph{HVdC} categories did not map directly to any of the themes in our taxonomy. This is the case with \emph{Misfortune/Fortune}, for example. We found that these types of events are usually reported as elements in dreams that were predominantly characterised by other themes, such as \emph{Life events}, \emph{People and relationships}, or \emph{Violence and death}, for example.

\begin{figure}[t!]
\centering
\includegraphics[width=0.99\textwidth]{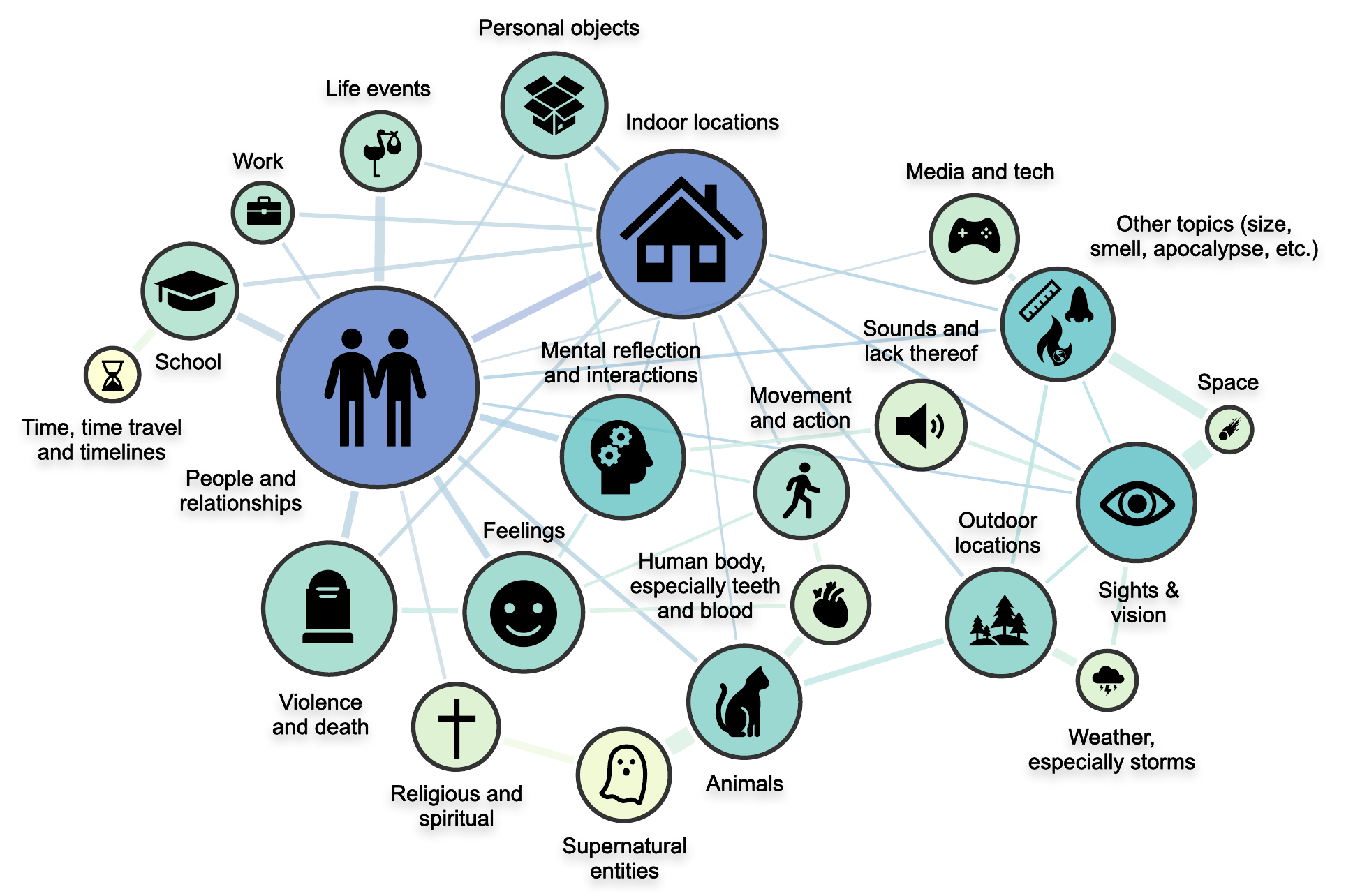}
\caption{\textbf{The taxonomy of dream themes that we automatically uncovered -- \emph{a map of dreams}\label{fig:taxonomy}}. Nodes represent various dream themes (i.e. groups of dream topics); node size corresponds to the frequency of the topics in our dream reports; node color represents the topic centrality -- the darker the color, the more central the topic; and the lighter the color, the more peripheral the topic; finally, edge thickness corresponds to the frequency of co-occurrence of the topics in our dream reports.}
\end{figure}

The emerging themes in the dream reports do not exist in isolation; they often co-occur in the same reports and jointly construct their narratives. Figure~\ref{fig:taxonomy} presents the backbone of the network of co-occurrence of these themes, where the connections between themes are weighted proportional to the number of dreams in which they co-occur. \emph{People and relationships} and \emph{Indoor locations} are the most frequent and central themes, occurring often with many other ones. Feelings are mainly associated with actions and relationships rather than objects or settings; conversely, sensory elements are more associated with settings, especially outdoor locations. Some of the central topics, such as relationships and emotions, are highly valued in rating scales by dream researchers. However, other themes, such as indoor/outdoor settings, are frequently omitted in \emph{HVdC} coding research. 

Besides the prominence of both types of settings, their proportions differ somewhat. Our data show a greater predominance of indoor settings, while \emph{HVdC} norms indicate lower levels of indoor settings: 49\% for males and 61\% for females \cite{hall1966content}. This difference could reflect our data, which includes two years of the COVID pandemic, whereas the comparison \emph{HVdC} studies do not.

 Our findings challenge traditional dream analysis literature that found males mostly dream of male characters.~\cite{domhoff2003scientific} The two most recurring topics in our data concern the presence of female characters in dreams, despite the young male dominance in the Reddit user base.~\cite{hargittai2020potential}
 A possible reason for this contradicting finding could have to do with \emph{HVdC} ratings not being exactly equivalent to ours. In our method, the strength of association of a dream with topics concerning male characters and indoor settings is stronger the more mentions of male characters and indoor settings appear in the report. Conversely, in \emph{HVdC}, a male character who is mentioned only once in the dream gets the same score as one who is referred to in every sentence of the report. Likewise, an indoor setting is scored in the same way regardless whether it is inferred from one mention in the dream or whether the dream account is largely a description of an indoor setting. In that sense, our methods allows for a richer characterisation and \emph{quantification} of dream content.

\begin{table}[!ht]
\centering
\footnotesize
\caption{Top 20 topics by the number of corresponding dream reports. \label{tab:topics}}
\begin{tabular}{p{.45\textwidth}p{.15\textwidth}p{.15\textwidth}}
\toprule
\textbf{Topic} & \textbf{\# dream reports} & \textbf{\# dreamers} \\ \midrule
lady, face, looked, head & 13020 & 11356 \\
dream, girl-dream, dreamt, ve & 6785 & 6337 \\
lights, sun, pitch-black, wearing & 4950 & 4488 \\
mall, restaurant, eating, ice-cream & 4660 & 4114 \\
bus, driving, cars, train-station & 4291 & 3966 \\
ex, years, dating, talk & 3786 & 3590 \\
kitten, lion, birds, owl & 3281 & 3060 \\
dreams-mean, staring, talk, people-dream & 3194 & 3018 \\
beach, ocean, swim, river & 3119 & 2900 \\
doors, house, rooms, mansion & 2984 & 2741 \\
death, died-dream, going-die, death-dream & 2825 & 2711 \\
classroom, teachers, principal, student & 2769 & 2565 \\
dream-dad, dad-dream, dream-mom, mom-dream & 2327 & 2197 \\
mum, aunt, mom-came, called-mom & 1893 & 1793 \\
felt-real, dream-felt, real-dream, real-like & 1840 & 1769 \\
dream-ends, story, endings, recurring-theme & 1677 & 1584 \\
know-make, sure-don, know-think, know-sure & 1649 & 1580 \\
school-dream, dream-school, college, dream-high & 1632 & 1553 \\
creature, bite, mouth, face & 1621 & 1555 \\
path, garden, hills, plants & 1595 & 1494 \\
\bottomrule
\end{tabular}
\end{table}

\begin{table}
\centering
\scriptsize  
\caption{\label{tab:topic_clusters}Dream themes and examples of their topics. For each theme, compatible categories from the Hall \& Van de Castle rating system are shown (\url{https://dreams.ucsc.edu/Coding/}). \cite{hall1966content} Rank represents the theme's/topic's overall prevalence across all dreams; the smaller the rank, the more prevalent the theme/topic.}
\begin{tabular}{p{.12\textwidth}p{.34\textwidth}p{.05\textwidth}p{.06\textwidth}p{.07\textwidth}p{.14\textwidth}}%
\toprule
\textbf{Theme} & \textbf{Top-3 topics under this theme} & \textbf{\# topics} & \textbf{\# dreams} & \textbf{\# dreamers} & \textbf{Hall/Van de Castle cat.}\\ \midrule



\begin{tabular}[c]{@{}l@{}}People and\\relationships\end{tabular} &  \begin{tabular}[c]{@{}l@{}}lady, face, looked, head\\      dream, girl-dream, dreamt, ve\\    ex, years, dating, talk\end{tabular} &   17 &  20786 &   17798 &  \begin{tabular}[c]{@{}l@{}} Characters;\\  Social Interactions \end{tabular}  \\  
\arrayrulecolor{black!20}\midrule

Animals & \begin{tabular}[c]{@{}l@{}}kitten, lion, birds, owl\\      spider, maggots, batman, thanos\\      snakes, alligator, turtle, bite\end{tabular} & 13 & 5546 & 5048 & \begin{tabular}[c]{@{}l@{}} Characters -> Animals \end{tabular}\\
\arrayrulecolor{black!20}\midrule

Supernatural entities & \begin{tabular}[c]{@{}l@{}}creature, bite, mouth, face\\      shadows, shadowy-figure, dark-figure, entity\\      zombies, zombie-apocalypse, outbreak, like-zombie\end{tabular} & 11 & 3436 & 3173 & \begin{tabular}[c]{@{}l@{}} Characters -> Imaginary \end{tabular} \\
\arrayrulecolor{black!20}\midrule

\begin{tabular}[c]{@{}l@{}}Religious and spiritual\end{tabular} & \begin{tabular}[c]{@{}l@{}}demons, devil, monster, demonic\\      church, cult, lot-people, dream-god\\      demon, devil, demons, angel\end{tabular} &  7 &  2859 &  2697 &  \begin{tabular}[c]{@{}l@{}}    Characters -> Imaginary \\  Objects -> Architecture \end{tabular}  \\ 
 \arrayrulecolor{black}\midrule



Indoor locations & \begin{tabular}[c]{@{}l@{}}mall, restaurant, eating, ice-cream\\      bus, driving, cars, train-station\\      doors, house, rooms, mansion\end{tabular} & 16 & 13890 & 11868 & \begin{tabular}[c]{@{}l@{}} Settings ->\\ Location
-> Indoor \end{tabular} \\ 
\arrayrulecolor{black!20}\midrule

Outdoor locations & \begin{tabular}[c]{@{}l@{}}beach, ocean, swim, river\\      path, garden, hills, plants\\      cave, tunnels, underground, tower\end{tabular} & 6 & 5020 & 4539 & \begin{tabular}[c]{@{}l@{}} Settings ->\\ Location
-> Outdoor \end{tabular} \\
\arrayrulecolor{black!20}\midrule

School & \begin{tabular}[c]{@{}l@{}}classroom, teachers, principal, student\\      school-dream, dream-school, college, dream-high\\      tests, failed, exam, professor\end{tabular} & 3 & 3639 & 3352 & \begin{tabular}[c]{@{}l@{}} Settings -> \\ Location \end{tabular}  \\
\arrayrulecolor{black!20}\midrule

Work & new-job, boss, jobs, shift & 1 & 1077 & 1024 & \begin{tabular}[c]{@{}l@{}}  Settings -> Familiar \end{tabular} \\
\arrayrulecolor{black!20}\midrule

\begin{tabular}[c]{@{}l@{}}Weather, especially \\storms \end{tabular} & \begin{tabular}[c]{@{}l@{}}rain, tornado, hurricane, started-raining\\      snow, snowing, cold, winter\\      tornado, storms, april, category\end{tabular} & 3 & 1037 & 985 & Settings \\ 
\arrayrulecolor{black}\bottomrule



\begin{tabular}[c]{@{}l@{}}Mental reflection\\and interactions\end{tabular} & \begin{tabular}[c]{@{}l@{}}know-make, sure-don, know-think, know-sure\\      yeah, like-wtf, hell, like-fuck\\      asked-doing, tell, help, begged\end{tabular} & 18 & 6808 & 6157  & \begin{tabular}[c]{@{}l@{}} Activities -> \\Thinking \end{tabular}
\\ 
\arrayrulecolor{black!20}\midrule

Sights and vision & \begin{tabular}[c]{@{}l@{}}lights, sun, pitch-black, wearing\\      reflection, looked-mirror, looking-mirror, mirrors\\      blinded, vision-blurry, blur, recite\end{tabular} & 11 & 6185 & 5548 & \begin{tabular}[c]{@{}l@{}} Activities -> \\ Visual; Auditory \end{tabular} \\
\arrayrulecolor{black!20}\midrule

Movement and action & \begin{tabular}[c]{@{}l@{}}left, wanted-leave, time-leave, wanted-home\\      continued-walking, continue, street, walk-home\\      run, started-running, run-like, sprinting\end{tabular} & 12 & 3457 & 3208 & \begin{tabular}[c]{@{}l@{}} Activities -> \\ Movement \end{tabular} \\
\arrayrulecolor{black!20}\midrule

\begin{tabular}[c]{@{}l@{}}Sounds and\\lack thereof\end{tabular} & \begin{tabular}[c]{@{}l@{}}singing, songs, lyrics, stage\\      voices, heard-voice, hear-voice, voice-head\\      footsteps, ghost, noises, ringing\end{tabular} & 7 & 3125 & 2892 & \begin{tabular}[c]{@{}l@{}} Activities -> \\ Auditory \end{tabular} \\
\arrayrulecolor{black}\midrule



Violence and death & \begin{tabular}[c]{@{}l@{}}death, died-dream, going-die, death-dream\\      pistol, shooting, shot-head, shotgun\\      police, officer, officers, kidnapped\end{tabular} & 15 & 8203 & 7523 & \begin{tabular}[c]{@{}l@{}} Social Interactions ->\\ Aggression \end{tabular}
\\ 
\arrayrulecolor{black!20}\midrule

Life events & \begin{tabular}[c]{@{}l@{}}birth, babies, pregnancy, newborn\\      party, invited, having-party, brother\\      giving-birth, dreamt, dream, twins\end{tabular} & 6 & 2230 & 2134 & \begin{tabular}[c]{@{}l@{}} Social Interactions ->\\ Friendliness \end{tabular} \\ 
\arrayrulecolor{black}\midrule



Personal objects & \begin{tabular}[c]{@{}l@{}}dressed, mask, naked, clothing\\      phones, ringing, check-phone, battery\\      pages, pen, ink, letters\end{tabular} & 17 & 4639 & 4135 & \begin{tabular}[c]{@{}l@{}} Objects \end{tabular} \\
\arrayrulecolor{black!20}\midrule

Media and tech & \begin{tabular}[c]{@{}l@{}}theater, anime, movie-like, tv\\      minecraft, games, vr, game-like\\      game-dream, dream-playing, vr, video-games\end{tabular} & 6 & 2970 & 2677 & \begin{tabular}[c]{@{}l@{}} Objects -> \\ Communication \end{tabular} \\
\arrayrulecolor{black!20}\midrule

\begin{tabular}[c]{@{}l@{}}Human body,\\esp.\ teeth and blood\end{tabular} & \begin{tabular}[c]{@{}l@{}}blood, skin, humanoid, arms\\      teeth, falling, tongue, pain\\      teeth, tooth, falling, gums\end{tabular} & 5 & 2163 & 2055 & \begin{tabular}[c]{@{}l@{}}  Objects -> \\ Body Parts \end{tabular} \\
\arrayrulecolor{black!20}\midrule

Space & \begin{tabular}[c]{@{}l@{}}sun, earth, eclipse, phases\\      space, space-ship, nasa, oxygen\\      meteors, earth, asteroid, coming\end{tabular} & 3 & 448 & 438 & \begin{tabular}[c]{@{}l@{}}  Objects -> \\ Nature \end{tabular} \\ 
\arrayrulecolor{black}\midrule



Feelings & \begin{tabular}[c]{@{}l@{}}felt-real, dream-felt, real-dream, real-like\\      woke-crying, started-crying, woke-tears, crying-dream\\      feel-pain, painful, pain-dream, felt-pain\end{tabular} & 16 & 6351 & 5907 & \begin{tabular}[c]{@{}l@{}} Emotions \end{tabular} \\
\arrayrulecolor{black}\midrule



\begin{tabular}[c]{@{}l@{}}Time, time travel\\and timelines\end{tabular} & \begin{tabular}[c]{@{}l@{}}timeline, time-travel, time-skip, like-time\\      time-travel, dream-world, universes, dream-time\\      noon, 00am, early-morning, evening\end{tabular} & 3 & 777 & 739 & \begin{tabular}[c]{@{}l@{}} Descriptive Elements -> \\ Temporal Scale \end{tabular} \\
\arrayrulecolor{black}\midrule


Other topics & \begin{tabular}[c]{@{}l@{}}dream-ends, story, endings, recurring-theme\\      huge, inches, like-size, big-small\\      pov, 3rd-person, person-perspective, person-view\end{tabular} & 21 & 5475 & 5018 & \begin{tabular}[c]{@{}l@{}} - \end{tabular} \\
\arrayrulecolor{black}\midrule

\end{tabular}
\end{table}

\begin{table}[!ht]
\centering
\caption{\label{tab:topics_by_dream_type} 
\textbf{Odds-ratio (OR) analysis for individual topics across different dream types.} The OR scores are equal to the probability of the topic appearing in dreams labeled with a given dream type (e.g., nightmare), divided by the same probability calculated on dreams that are \emph{not} labeled with that type. Scores higher/lower than one indicate that dreams of the given type feature the topic more/less often than other dreams. Unlike in Figure \ref{fig:specific_themes_dream_types}, in which we reported OR across themes; in this table, we report OR across individual topics to finely characterize a certain dream type by a list of specific topics appearing in it.
}
\small
  \begin{minipage}{.50\textwidth}
    
    \begin{tabular}{p{.8\textwidth}p{.08\textwidth}}
      \multicolumn{1}{c}{\textbf{Topics}} &
      \multicolumn{1}{c}{\textbf{OR}} \\
      \toprule
      \multicolumn{2}{c}{\textbf{Nightmares}}\\
      \midrule
        shadows, shadowy-figure, dark-figure, entity & 3.986 \\
        rape, sexually-assaulted, abuse, nightmares & 3.184 \\
        wasn-scary, scary-just, frightening, scariest & 3.017 \\
        really-creepy, weird-creepy, creepiest, sounds & 2.880 \\
        tw, warning, nsfw, sexual-assault & 2.799 \\
        violent, disturbing, dreams-past, weird-dreams & 2.476 \\
        demons, devil, monster, demonic & 2.455 \\
        escape, escaped, way-escape, managed-escape & 2.421 \\
        911, dial, emergency, ambulance & 2.405 \\
        dolls, porcelain, voodoo, haunted & 2.267 \\
        panic, panicking, panicked, started-panic & 2.26 \\
        footsteps, ghost, noises, ringing & 2.235 \\
        feel-pain, painful, pain-dream, felt-pain & 2.207 \\
        blood, like-blood, covered-blood, splatter & 2.204 \\
        paralyzed, able, speak, couldn-speak & 2.109 \\
        lungs, couldn-breathe, like-couldn, heart & 2.012 \\
        serial-killer, dream-killing, killed-dream, killers & 1.956 \\
        blinded, vision-blurry, blur, recite & 1.889 \\
        demon, devil, demons, angel & 1.886 \\
        feel-right, felt-wrong, wasn-right, feel & 1.771 \\
        & \\
        \toprule
      \multicolumn{2}{c}{\textbf{Lucid dreams}}\\
      \midrule
        control, couldn-control, control-body, hopeless & 4.119 \\
        universes, alternate-reality, different-dimension, travel & 2.82 \\
        reflection, looked-mirror, looking-mirror, mirrors & 2.702 \\
        felt-real, dream-felt, real-dream, real-like & 2.196 \\
        paralyzed, able, speak, couldn-speak & 2.052 \\
        falling, hit-ground, fall-ground, let-fall & 1.979 \\
        know, okay, yeah, know-going & 1.855 \\
        voices, heard-voice, hear-voice, voice-head & 1.627 \\
        don-want, needed, doing-don, remember-doing & 1.535 \\
        demon, devil, demons, angel & 1.355 \\
        flying, airport, helicopter, planes & 1.355 \\
        & \\
      \bottomrule
    \end{tabular}
  \end{minipage}%
  \begin{minipage}{.50\textwidth}
    \begin{tabular}{p{.8\textwidth}p{.08\textwidth}}
      \multicolumn{1}{c}{\textbf{Topics}} &
      \multicolumn{1}{c}{\textbf{OR}} \\
      \toprule
      \multicolumn{2}{c}{\textbf{Vivid dreams}}\\
      \midrule
        felt-real, dream-felt, real-dream, real-like & 2.104 \\
        religious, atheist, catholic, supernatural & 1.722 \\
        feel-right, felt-wrong, wasn-right, feel & 1.649 \\
        apocalyptic, dream-end, world-dreams, earth & 1.614 \\
        feel-pain, painful, pain-dream, felt-pain & 1.500 \\
        know-knew, knew-didn, knew-don, known & 1.473 \\
        birth, babies, pregnancy, newborn & 1.456 \\
        aliens, invasion, grey, race & 1.445 \\
        path, garden, hills, plants & 1.435 \\
        reflection, looked-mirror, looking-mirror, mirrors & 1.419 \\
        giving-birth, dreamt, dream, twins & 1.356 \\
        4nuke, war, fires, missile & 1.313 \\
        & \\
      \toprule
      \multicolumn{2}{c}{\textbf{Recurring dreams}}\\
      \midrule
        cheating, having-dreams, dream-boyfriend, relationship & 6.349 \\
        ex-boyfriend, dreams, having-dreams, relationship & 4.021 \\
        teeth, tooth, falling, gums & 3.760 \\
        school-dream, dream-school, college, dream-high & 3.084 \\
        house-dream, dream-house, apartment, childhood-home & 2.966 \\
        ex, years, dating, talk & 2.462 \\
        dreams-mean, staring, talk, people-dream & 2.087 \\
        dream-ends, story, endings, recurring-theme & 2.047 \\
        dream, girl-dream, dreamt, ve & 1.974 \\
        apocalyptic, dream-end, world-dreams, earth & 1.912 \\
        time-travel, dream-world, universes, dream-time & 1.857 \\
        crush, falling-love, fell-love, fall-love & 1.799 \\
        teeth, falling, tongue, pain & 1.788 \\
        rape, sexually-assaulted, abuse, nightmares & 1.679 \\
        paralyzed, able, speak, couldn-speak & 1.658 \\
        flying, airport, helicopter, planes & 1.625 \\
        sex-dream, sexual, wet-dream, dream-sex & 1.624 \\
        serial-killer, dream-killing, killed-dream, killers & 1.556 \\
        room-starts, walking, family, house-starts & 1.517 \\
        doors, house, rooms, mansion & 1.465 \\
      \bottomrule
    \end{tabular}
  \end{minipage}%
\end{table}

\begin{figure}[!ht]
\centering
\includegraphics[width=0.99\textwidth]{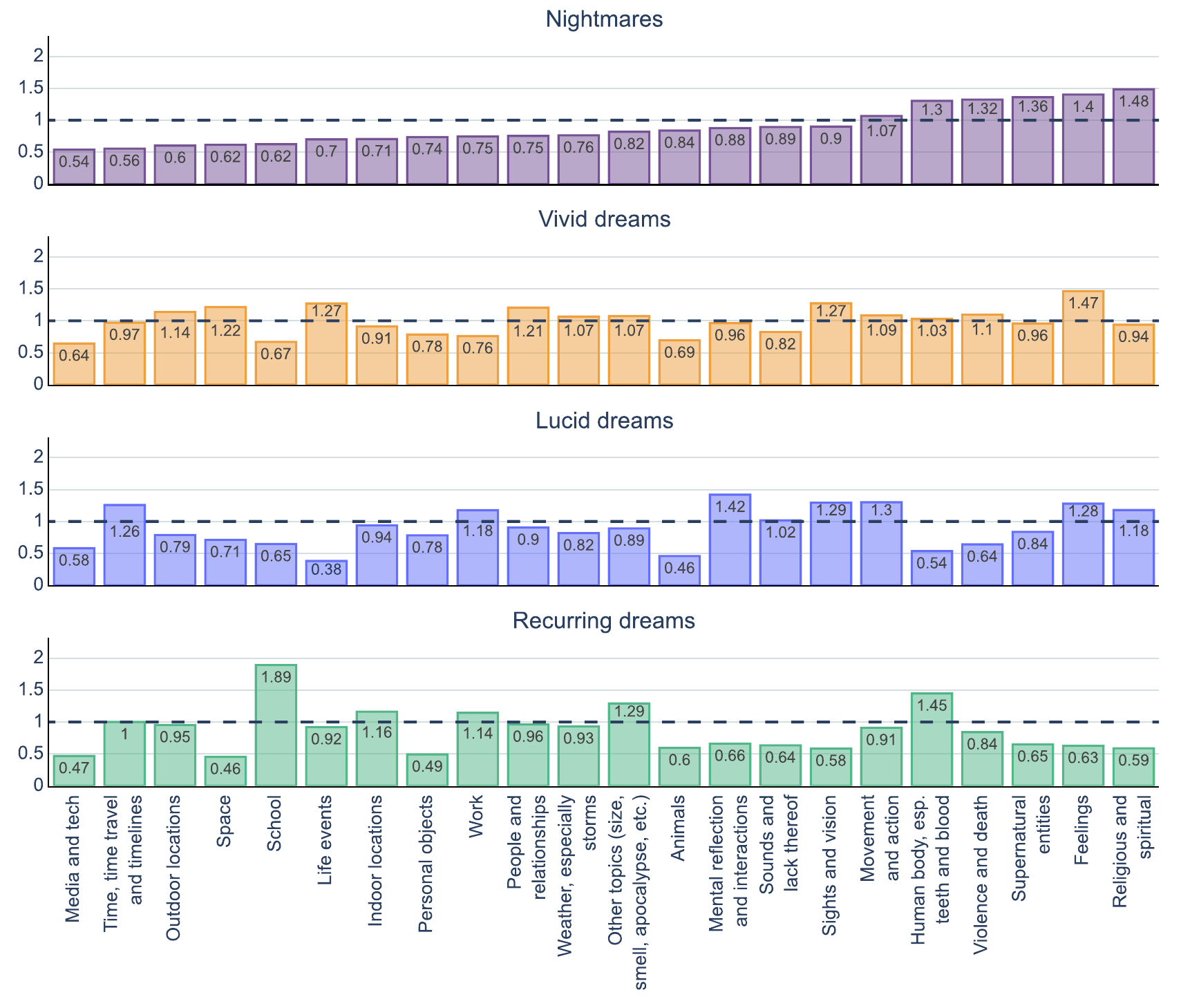}
\caption{\textbf{Odds-Ratio analysis of dream themes in various dream types.} This graph represents the probability of specific themes (x-axis) appearing in a given dream type (e.g., nightmare), relative to their appearance in all other dream types. Values above or below 1 on the y-axis signify that these themes occur more or less frequently, respectively, in the identified dream type compared to the aggregate of other dream types.\label{fig:specific_themes_dream_types}}

\end{figure}

\subsection*{Topics and themes by dream type}\label{sec:specific_dreams}

Using odds ratios (see Methods), we uncovered that certain topics (Table \ref{tab:topics_by_dream_type}) and themes (Figure \ref{fig:specific_themes_dream_types}) appeared more frequently in dreams of specific types.

\subsubsection*{Topics and themes in Nightmares}\label{sec:nightmare_specific}
Results in Table \ref{tab:topics_by_dream_type} revealed among the top topics specific for nightmares, keywords such as \emph{shadows}, \emph{rape}, \emph{sexual-assault}, \emph{scary}, \emph{creepy}, \emph{violence}, \emph{demons}, \emph{911}, and \emph{blood}.
In terms of themes, \emph{Religious and spiritual} is the most prominent in nightmares, followed by \emph{Feelings}, \emph{Supernatural entities}, and \emph{Violence and death}, while the least prominent ones were \emph{Media and tech} and \emph{Time, time travel and timelines}.

\subsubsection*{Topics and themes in Vivid dreams}\label{sec:vivid_specific}
Results presented in Table \ref{tab:topics_by_dream_type} reveal 
among topics specific for vivid dreams, keywords such as 
\emph{felt-real}, \emph{religious},  \emph{felt-right}, \emph{felt-wrong}, \emph{apocalyptic}, \emph{felt-pain}, \emph{baby birth and pregnancy}, \emph{aliens}, \emph{mirrors}, and \emph{nuclear war}. This has translated into the most prominent themes being \emph{Feelings}, followed by \emph{Sights and vision} and \emph{Life events}.

\subsubsection*{Topics and themes in Lucid dreams}\label{sec:lucid_specific}
Results presented in Table \ref{tab:topics_by_dream_type} reveal 
among topics specific for lucid dreams, keywords such as
\emph{control}, \emph{alternate-reality}, and \emph{reflection and mirrors}, \emph{felt-real}, \emph{couldn-speak}, \emph{falling}, \emph{heard-voices} and \emph{demon}. When considering themes, \emph{Mental reflection and interactions}, followed by \emph{Movement and action}, \emph{Sights and vision}, and \emph{Feelings} were the most prominent.

\subsubsection*{Topics and themes in Recurring dreams}\label{sec:recurring_specific}
Results in Table \ref{tab:topics_by_dream_type} revealed among the top topics specific for recurring dreams, keywords such as \emph{cheating}, \emph{ex}, \emph{teeth}, \emph{school},  \emph{house and apartment},  \emph{time-travel}, and  \emph{sex-dreams}.
Looking at the themes, we found \emph{School}, \emph{Human body, especially teeth and blood}, and \emph{Other topics (size, smell, apocalypse, etc.)} being those that characterize recurring dreams. 

\subsection*{Topics and themes through time}
\label{sec:temporal}
\begin{figure}[!th]
\centering
\includegraphics[width=0.99\textwidth]{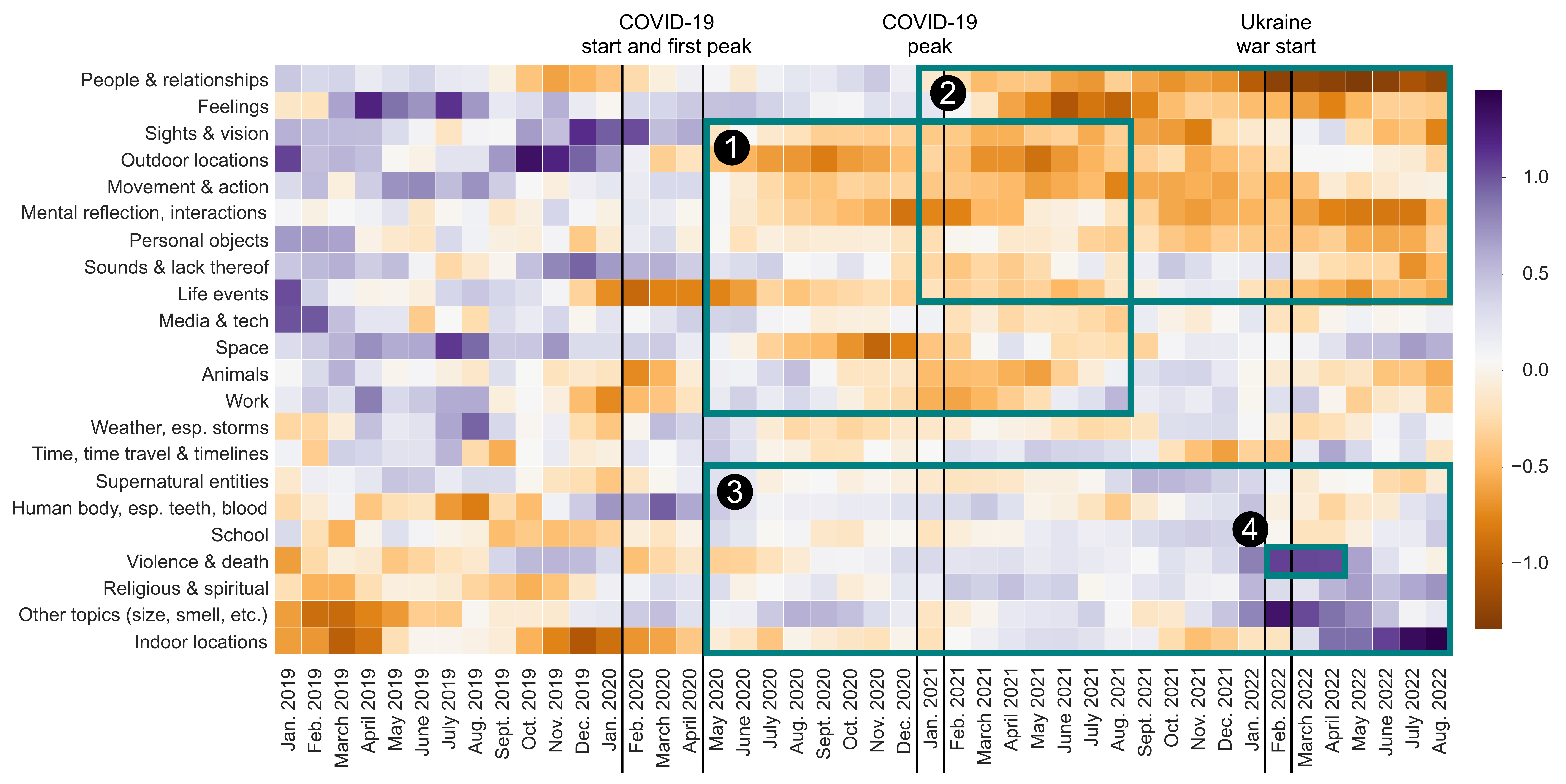}
\caption{\textbf{Temporal evolution of themes.} The value of each cell is proportional to the average importance of a theme $t$ across all the dream reports posted on a given month $m$ ($I_{t,m}$ in Formula~\ref{eq:topic_importance_avg}). The importance of a theme in a dream is defined as the fraction of sentences in the dream report that belong to that theme ($I_{t,d}$ in Formula~\ref{eq:topic_importance}). Values are standardized within each theme ($z_{I_{t,m}}$ in Formula~\ref{eq:zscore}) so that positive (negative) values encode values of importance that are higher (lower) than the global average for that theme. Last, we smoothed values over a window length of 5 months to reduce noise ($\widetilde{z}_{I_{t,m}}$ in Formula~\ref{eq:zscore_smooth}).  During period 1) themes of \emph{Sight and vision}, \emph{Outdoor locations}, \emph{Movement and action}, \emph{Mental reflections}, \emph{Life Events}, \emph{Space}, and \emph{Work} plummeted, while in period 2) also the themes of \emph{People and relationships} and \emph{Feelings} dropped. During period 3) we see an increase in the themes such as \emph{Supernatural entities}, \emph{Human body, especially teeth and blood}, \emph{Religious and spiritual}, \emph{Indoor Locations}, and \emph{Violence and death}. The last topic peaked around the time of the start of the war in Ukraine (shown by the box 4)).
\label{fig:temporal}
}
\end{figure}

Finally, we studied the evolution of topics and themes over time. We focused on the period with at least 300 monthly dream reports, namely from January 2019 to September 2022. We found that soon after COVID-19 started, there was a gradual collective shift in the content of dreams from \rdreams (Figure~\ref{fig:temporal}). 

At the very beginning of the COVID-19 outbreak (February-March 2020), and even more so after the first peak of recorded deaths (April 2020), people gradually dreamt less of \emph{Sight and vision}, \emph{Outdoor locations}, \emph{Movement and action}, and \emph{Mental reflections and interactions}, while they dreamt more of \emph{Religious and spiritual} figures, \emph{Indoor locations}, and \emph{Human body, especially teeth and blood}. An example dream from this period talks about \emph{``spitting teeth and cornea onto the palm.''} We found a sharp decrease in the frequency of \emph{Life events} topics in February 2020, while in March 2020 there we recorded a peak of mentions of \emph{Human body, especially teeth and blood}, which are predominantly found in nightmares, as our previous analysis showed. These trends continued throughout the time of the COVID-19 second death peak (January 2021), from when we also detect a stark decrease in dreams of other \emph{People and relationships}, \emph{Feelings}, and, for a while, of \emph{Animals} and \emph{Work}.

\begin{figure}[!th]
\centering
\includegraphics[width=0.99\textwidth]{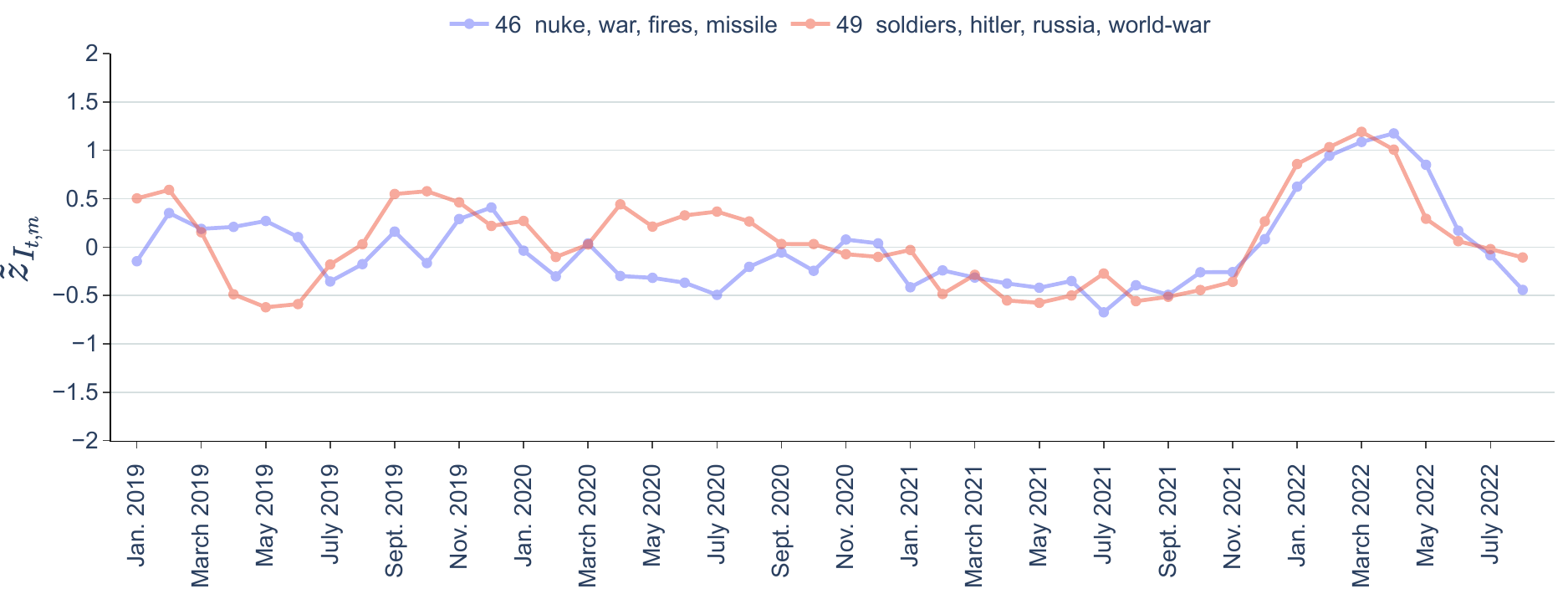}
\caption{\textbf{Temporal evolution of two selected topics: \emph{soldiers, hitler, russia, world-war} and \emph{nuke, war, fires, missile}.} Smoothed z-scores $\widetilde{z}_{I_{t,m}}$ ($y$-axis)  of a topic's importance at a given month ($x$-axis) are shown for the period from January 2019 to September 2022. The values $\widetilde{z}_{I_{t,m}}$ are proportional to the average importance of the topic $t$ across all the dream reports posted on a given month $m$ ($I_{t,m}$ in Formula~\ref{eq:topic_importance_avg}). The importance of a topic in a dream is defined as the fraction of sentences in the dream report that belong to that topic ($I_{t,d}$ in Formula~\ref{eq:topic_importance}). Values are standardized within each topic ($z_{I_{t,m}}$ in Formula~\ref{eq:zscore}) so that positive (negative) values encode values of importance that are higher (lower) than the global average for that topic. Last, we smoothed values over a window length of 5 months to reduce noise ($\widetilde{z}_{I_{t,m}}$ in Formula~\ref{eq:zscore_smooth}). Both topics peaked around the time when the conflict in Ukraine started.\label{fig:temporal_topics} }
\end{figure}

Finally, the start of the war in Ukraine (February 2022), is associated with a strong transition on the content of people's dreams towards violent topics. We found a sharp increase in topics from the themes of \emph{Violence and death}, and \emph{Other topics (size, smell, apocalypse, etc.)}. Example dreams from the former group talk about \emph{``being a murderer,''},\emph{``tooth falling out,''} and \emph{``getting shot in the head.''} Example dreams from the second group talked about \emph{``temple spirit monster,''} and \emph{``being attacked by an opossum.''} The changes in response to external events were also evident at the level of individual topics, such as those about soldiers and nuclear war, which both peaked after the war in Ukraine started (Figure~\ref{fig:temporal_topics}).

\section*{Discussion}\label{sec:discussion}

The current study advances the field of dream science by implementing a new methodology to study dreams in a more objective and ecological manner and also showcasing how this method can generate new insights into dreams. 

\subsection*{Reddit as a source of dreams}
Prior dream research has relied heavily on traditional laboratory, survey, and diary methods. Laboratory studies benefit from monitoring participants with PSG and waking them directly from REM sleep,\cite{siclari2013assessing} which is known to increase dream recall drastically.\cite{nemeth2022route} However, these dreams are not representative of natural dream content, as they are highly influenced by the laboratory setting.\cite{picard2021dreaming} Survey studies avoid this contextual bias on dream content, but often ask of a ``recent'' dream, which might be days or weeks prior to the survey response, thus suffering from memory distortion. Lastly, diary studies track dream content in a participant sample through morning diaries.\cite{schredl2002questionnaires} While these studies benefit from having dream content collected from ecological settings and with less retrospective bias, these studies are often limited by small sample sizes due to the high burden of participation. In the current study, we extracted morning dream reports from social media, thus capturing dreams in an ecological setting and also at a much larger scale. While other studies have investigated dedicated online dream forums,\cite{sanz2018experience} Reddit is one of the most popular social media sites and its usage continues to grow at higher rates than specialized forums, suggesting that the current approach might continue to be a source of population dream content for scientific analysis.

\subsection*{Unsupervised generation of dream themes}
In using this ecological data source to generate common dream themes, our results complement previous studies using more traditional survey methods. Though it is widely accepted that dream content varies based on individual personality and cultural differences, previous research suggests there might also be thematic ``universals'' that appear in a disproportionately high amount of dreams. Universal dream themes are typically quantified using surveys with predetermined thematic content developed by the researcher,\cite{nielsen2003typical,yu2012pornography} which are biased towards existing knowledge of dreams. In the current study, our unsupervised approach to developing common dream themes confirms some previously developed themes while also offering more specificity within them. For example previous survey studies using the Typical Dreams Questionnaire\cite{nielsen2003typical} or Dream Motif Scale\cite{yu2012pornography} have identified common dream themes of failure, paranoia, snakes/insects and animal symbolism, alien life, fighting, and sex. Our results identified similar themes, while offering a finer-grained view with the subtopics that formed each theme. For example, we observed a popular theme of animals, and animal subtopics included what might be positive animal dreams (kitten, birds) and negative animal dreams (spider, maggots; snake, bite). It is difficult to compare the ranking of our dream themes with prior work, since prior work is heavily-dependent on the method of data collection.\cite{mathes2014frequency} A notable advance of our approach is the ability to ``map out'' the relationship between dream themes presence. The co-occurence of common themes has not been studied extensively before, and future work comparing the co-occurence of waking and dreaming themes might help to uncover what is truly unique about dream content. 

\subsection*{Phenomenology of dream subtypes} 
Dreams are highly varied experiences, and have long been grouped into subclasses or types of dreams (e.g., nightmares, lucid dreams). However, these dreams are often defined by a single
feature (e.g., nightmares as intensely negative dreams) and the phenomenological variety within each subtype is not well understood. The present results offer new insights into the consistent-yet-variable content of nightmares, lucid dreams, vivid dreams, and recurring dreams, some of which have important clinical implications. 

Nightmares are defined as intensely negative dreams, sometimes with a secondary requirement that they result in a direct awakening, \cite{gieselmann2019aetiology} and have immense clinical relevance in PTSD \cite{campbell2016nightmares} and other psychiatric diagnoses. \cite{sheaves2022nightmares} Our observation of keywords relating to sexual assault highlight prior observations of uniquely episodic event replay in PTSD patients, \cite{phelps2008understanding} including survivors of sexual assault.\cite{krakow2002nightmare} The additional observation of increased themes of Feeling and Violence suggest that the episodic replays are highly emotional recreations of violent events, and the inclusion of many escape-related words highlights the helplessness felt by many recurrent nightmare sufferers. \cite{rousseau2018mechanisms} For nightmare sufferers, the negative affect during dreams \cite{mallett2021relationship} or the dream recall during the day might increase negative symptoms of other co-morbid diagnoses (e.g., anxiety). \cite{blagrove2004relationship} Lastly, the heightened presence of supernatural entities in nightmares might relate to the common state of sleep paralysis, an under-studied and cross-cultural phenomenon that occurs during sleep-wake transitions and frequently involves a feeling of helplessness amidst a hallucinated ``demon'' or otherwise frightening figure.\cite{denis2018systematic} 

Vivid dreams are highly realistic (or similarly, well-remembered) dreams. Our unsupervised approach suggests that vivid dreams are not only realistic (e.g., keywords felt-real, real-like), but also that these dreams often contain major life events, strong emotions, and supernatural/religious experiences. Vivid dreams included births and pregnancies, missiles and war, apocalyptic events, and alien invasions. This dream subtype overlaps almost directly with a class of dreams referred to as ``big dreams.'' \cite{bulkeley2011big} Big dreams occur rarely, but when they do, they are highly meaningful experiences that make a significant and long-lasting impression on waking life. Thus, our analysis of vivid dreams might be representative of these dreams, given that they consisted of major life events and religious experiences that likely influenced future thinking. Interestingly, the presence of an Alien invasion theme in vivid/realistic dreams suggests that prior reports of UFO abductions might result from cases of dream-reality confusion, \cite{wamsley2014delusional} where a dreamt abduction is misinterpreted as a memory from waking life. \cite{holden2002alien} 

Lucid dreams are defined as those that include awareness of the dream while still dreaming, \cite{mallett2021exploring} sometimes with a secondary requirement of having control over the dream. \cite{windt2018spontaneous} The present unsupervised analysis was consistent with these defining features, with common themes and keywords in lucid dreams such as mental reflection and control. Additionally, our results confirm more recent preliminary findings about lucid dream content, such as frequent episodes of flying \cite{picard2020flying} and an overlap with sleep paralysis. \cite{mainieri2021sleep} Though lucid dreams are generally regarded as more positive in valence than non-lucid dreams, \cite{mallett2020partial,schredl2022differences,voss2013measuring} there are more recent reports of extremely negative lucid dreams, or lucid nightmares. \cite{mallett2022benefits,schredl2020lucid,stumbrys2018lucid}  Our results offer a cohesive explanation for these differential findings, in that we observed a general heightened realness and emotion in lucid dreams (themes of Feeling and Sights and visions and keywords of felt-real) without attachment to positive or negative valence. Our recent findings, focused on a different subreddit (r/LucidDreaming) suggest that positively-valenced lucid dreams are more likely to occur when dream control is involved, \cite{mallett2022benefits} and the current results highlight the importance of focusing future clinical applications of lucid dreaming on the dream control rather than simply awareness of the dream (see \cite{ouchene2023effectiveness} for a review of the clinical efficacy of lucid dreams to treat nightmares).

A recurrent dream is one that is experienced repeatedly, and these have been estimated to occur at least once in roughly 75\% of the population.\cite{vallat2018sleep,zadra1996recurrent} It is likely that the many themes and keywords we observed in recurrent dreams are related to waking anxieties or worries. The limited amount of prior work on recurrent dream phenomenology suggests that recurrent dreams are primarily related to waking anxieties \cite{robbins1992comparison} and other negative content, \cite{weinstein2018linking} and also increase in frequency during periods of stress. \cite{duke2002ordinary,zadra1996recurrent} Our analysis extends these findings by observing more specific anxieties in recurrent dream content. The top two themes were related to relationships, particularly negative aspects of relationships (i.e., cheating, ex-partners), and other common themes were even still related to relationships (e.g., sex, dating, crush). Other recurrent dream themes were explicitly negative, and at times ultraviolent; themes regarding serial killers, paralysis, sexual assault, and the apocalypse suggest that recurrent dreams are far more negative than positive or neutral. We suspect that many of these recurrent dreams would also be classified as nightmares (see statistics in Figure SI1), and thus our analysis might help future work in the prediction/monitoring of nightmares via the inclusion of recurrent themes.

\subsection*{Impact of major collective events on dreams}
Previous research has revealed that major personal and cultural events might influence dream content. For example, an increase in nightmares was observed after the terrorist attacks of September 11th, 2001 \cite{propper2007television} and during the COVID-19 pandemic. \cite{gorgoni2022dreaming,margherita2022observatory} Dreams during the COVID-19 pandemic have also been shown to include pandemic-related content. The present results expand on these prior findings by offering a finer-grained view into the topical and temporal impact of major events on dreams. Rather than a categorical increase in nightmares or pandemic-related content, our analysis allowed us to identify specific sub-themes of pandemic-related topics and how they change over time. During the COVID-19 pandemic, population dreams transitioned from outdoor to indoor locations and decreased in social interactions, as did our waking experiences during the pandemic. Interestingly, these two effects had qualitatively different time courses, where the location change of dreams was longer-lasting and more persistent than social changes. This might reflect the mass adoption of technical communications (e.g., Zoom gatherings) that occurred while people were still mostly indoors. These effects are consistent with prior work showing a continuity between wake and dream content  \cite{fogli2020our,schredl2003continuity} and future work might evaluate how subtopics contribute uniquely to this continuity. \cite{schredl2003continuity} These results also contribute to hypothesized dream functions, such as the drop in social content contradicting the Social Simulation Theory that predicts a compensatory effect of social activity in dreams. \cite{tuominen2022no} We also observed more mental reflections in dreams, which might reflect a ``cognitive continuity'' hypothesis \cite{domhoff2017invasion} if the population was more reflective during the increased loneliness during the pandemic.\cite{kauhanen2022systematic} With the state of mental reflections during waking being less established than location and social changes, our finding suggests that it might be possible in the future to infer waking cognitive changes based on those observed in dream content. Alternatively, the rise in mental reflection during dreams might be representative of the increase in lucid dream frequency during the pandemic.\cite{kelly2022lucid} 

While dreams during the COVID-19 pandemic have been investigated at great length \cite{gorgoni2022dreaming,margherita2022observatory,vscepanovic2022epidemic,solomonova2021stuck}, much less work has been dedicated to observing the influence of the Russo-Ukrainian war on sleep and dream patterns. We observed a population increase in negative war-related topics (e.g., violence, death) after the start of the war, which is notable because our social media sample is expected to be primarily American users and thus not those who were directly exposed to the war. The association between the war and population dreams could be a result of American media exposure (see also \cite{propper2007television}, highlighting a cognitive continuity between dreams and wake, where it is not daily activities per se, but the thought processes and internal imagery that predicts dream content). The negative content of population dreams during the war has important implications, given recent findings that negative dreams, including specifically dreams of death, are predictive of next-day negative affect \cite{mallett2021relationship} and nightmares have extensive mental health implications.\cite{sheaves2022nightmares} 

\subsection*{Limitations}\label{sec:limitations}

The first limitation of our work has to do with potential biases in the dream reports that we analysed. The first bias is that Reddit users report dreams that they have recalled spontaneously. However, according to the salience hypothesis, dream content that is vivid, new, intense, or strange, is more easily remembered.\cite{watson2003dream} Moreover, personality traits also affect dream recall, so that people with higher openness to experience, and who are prone to imagination and fantasy, more often recall their dreams.\cite{watson2003dream} Finally, the users of Reddit are not a representative sample of the general population; they are known to be more male, young, educated, and urbanites.\cite{hargittai2020potential}

The second limitation is about the elements from dream reports not reflecting the dream content. Given the free-form social media format, the \rdreams \space users would not always describe only their dream report, but sometimes they would include some contextual information, for example recounting how they felt when they were woken up from the dream, or how the people they dreamt of are related to them in real life. For example, a Reddit user might drop in their dream report a sentence like \emph{``It was about 5 am when I woke up from this dream...''} Such contextual information is not a direct description of the dream content, yet it often helps to qualify it and it is therefore useful to our analysis. Dream researchers analysing a small number of dreams could read through each report and manually remove such instances to focus on dream content only. Given the automatic topic extraction method that we employed, such a data cleaning step was not possible. Some common categories of contextual information that were unrelated to the dream experience (e.g., the author explaining when they last time met the person they dreamt of, or what time they woke up from the dream) emerged as independent topics in our topic analysis and we removed them (see Methods, Clustering Topics).

The third limitation of our work is shared with other content analysis methods.\cite{schredl2010characteristics} We inevitably lose some dream information that could not be captured by the topics. This also means that our approach cannot represent subtle aspects of the individuality of each dreamer.

\subsection*{Implications}\label{sec:implications}

Our work has three main implications:

\paragraph{Developing an unsupervised dream content analysis method.} The development of dream scoring systems has historically favored certain aspects of dream reports over others, based on assumptions about what are the most emotionally and socially important dimensions in dream analysis, rather than considering all potential topics or types of human experiences. Previous attempts to apply machine learning to dream content analysis have either replicated existing scoring systems\cite{fogli2020our} or focused on specific types of dream content (e.g., symptoms\cite{vscepanovic2022epidemic}). In contrast, our study used a deep learning approach that prioritizes no particular topic. By analyzing dream content from an exclusively linguistic standpoint, we were able to discover and quantify themes that have not been previously considered.

\paragraph{Uncovering the first ecological taxonomy of dream topics.} Our results demonstrate that many of the themes we discovered align with those captured by existing scoring systems, particularly the Hall and Van de Castle scale, as it includes includes topics such as interpersonal relationships, emotions, and friendly and violent interactions. However, our approach also revealed differences in the frequency of certain topics, such as a significant category of weather-related topics that have not been previously considered in any dream scoring system. 
The significance of weather as a conversational topic varies, being mentioned as a safe subject for small talk or as the subject of jokes regarding dull conversations. Nevertheless, considering the evolutionary perspective, weather played a crucial role during the era of limited shelters and the absence of temperature controls, which shaped human instincts. Hence, weather likely holds a deep-rooted importance for our well-being and survival. To sum up, our findings support the continued use of traditional scoring systems in clinical psychology research and psychotherapy, where there is a strong rationale for focusing on the more emotional and social content of dreams. At the same time, our results suggest that AI tools can provide a more detailed and nuanced understanding of dream content, and may be mature enough to support dream analysts in their work.

\paragraph{Collaborating with AI in scientific discovery.} The AI's ability to categorize dream content in ways unknown to human researchers is reminiscent of the scenario when chess and go-playing programs began surpassing human players. These programs didn't merely excel at the strategies employed by humans, instead, they developed unique strategies that had never been observed by humans before. It was assumed that humans approached both games based on evolutionary instincts developed for social interactions, while the AI adopted a more objective perspective, solely focusing on the game rules without any assumptions influenced by human endeavors. Similarly, AI comprehends texts based solely on their intrinsic content, without filtering them through instinctual categories primarily designed for interactions during waking states.

\section*{Methods}\label{sec:methods}

\subsection*{Data pre-processing}\label{sec:data_preprocessing}

For the purpose of topical analysis, we employed the \texttt{en\_core\_web\_sm} model from \emph{Spacy} (\url{spacy.io}) to segment dream reports into individual sentences. The corpus consisted of 44,213 dream reports, for a total of 761,619 sentences. On average, each dream report contains 17 sentences and 290 words. Notably, recurring dreams tended to be shorter, with 14 sentences and 253 words, while lucid dreams were longer, with 27 sentences and 465 words (Table~\ref{tab:stats}). It is worth mentioning that the distribution of dream reports was not concentrated among a few individuals; the majority of users shared a single dream report, whereas only a small group of frequent users contributed more than 30 dream reports (Figure~\ref{fig:dreams_per_user}).

We discarded the top frequent 10,000 sentences with the fewest characters. These sentences typically contained non-dream related content, such as greetings to the reader (hello, hi, etc.), sentences consisting solely of special characters, and similar instances. This step of removing a significant number of sentences that were not relevant to the dream report itself ensured that our subsequent labeling process for non-dream topics was sufficient to preserve predominantly dream topics.

\subsection*{Topic Modelling}\label{sec:topicmodelling}
Our topic modelling procedure consisted of the following five steps: i) extracting topics, ii) grouping topics into themes, iii) building dream topics taxonomy, iv) finding topics and themes by dream type, and v) finding topics and themes through time.

\subsubsection*{Extracting topics}\label{sec:topicmodelling:extract-bertopic}
BERTopic~\cite{grootendorst2022bertopic} is largely based on a neural model that is designed to identify latent topics within document collections. Unlike conventional topic modeling techniques such as Latent Dirichlet Allocation, BERTopic leverages semantic information by utilizing embeddings as an initial step to cluster documents into semantically cohesive topics. Each discovered topic in BERTopic is described by a list of 10 \emph{topic words}, which are the most distinctive words associated with that particular topic. The topics are numbered based on their frequency rank, indicating their prevalence within the corpus.

In its default configuration, BERTopic assigns a single topic to each document. However, our manual inspection of the dream reports revealed that most dreams cannot be adequately characterized by a single topic. Instead, they often encompass multiple topics, such as the dream's location, the people involved, and the emotions experienced by the dreamer. Additionally, dreams are known for combining various elements from waking experiences, resulting in a sense of bizarreness. To address this, our initial alternative was to modify BERTopic to associate a distribution of topics with each dream report. We tested this approach by allowing up to ten topics per report. Subsequently, we applied a threshold to the probabilities associated with each topic to identify the relevant topics for each report. However, we encountered two issues with this method. Firstly, due to the substantial variability in the length of dream reports, we often missed relevant topics in longer dreams. Secondly, even with varying the threshold, over 55\% of dreams ended up being associated with no topic, resulting into a considerable loss of data. To mitigate this issue, we opted to consider individual dream sentences as input documents for BERTopic. Applying BERTopic at sentence-level is a practice recommended by the BERTopic authors on the official website of the tool. Such a solution enabled us to associate over 88\% of dream reports with at least one topic.

To establish robust topic representations, we configured the hyperparameters of BERTopic. We set the minimum frequency threshold (\texttt{min\_df}) to 10, ensuring that a word appears in at least 10 sentences before it is considered for inclusion in the topic representation. This criterion helped to ensure that topics were formed based on words with a reasonable level of occurrence within the dream reports. Additionally, we employed the Maximal Marginal Relevance (MMR) algorithm with a diversity parameter set to 0.4. The MMR algorithm was utilized to enhance the diversity of the topic words, preventing dreams from being described solely by near-synonyms. By incorporating this diversity measure, the resulting topics encompassed a wider range of relevant terms, capturing distinct aspects and avoiding redundant or similar descriptions within the topic representation.

Our model successfully extracted a total of $288$ topics which included both dreaming and non-dreaming subjects. An exhaustive manual inspection of the topic words demonstrated a significant degree of semantic coherence across the majority of topics: for most topics, the top four topic words provided sufficient information to understand the essence of the topic. To ensure topic specificity, we excluded all instances of the terms ``dream'' and ``dreams'' from the list of topic words associated with each topic. Furthermore, we aggregated topics related to multiple sentences within the same dream, consolidating them into a comprehensive list. This approach allowed us to capture the overarching themes and content associated with individual dreams more effectively.

\subsubsection*{Grouping topics into themes}\label{sec:topic-clustering}
To provide a concise overview of the $288$ identified topics, we used clustering techniques to group them into broader, yet semantically-coherent themes. To achieve this, we used the Sentence Transformer model \emph{all-mpnet-base-v2}~\cite{devlin2018bert} to project each topic word into a 768-dimensional embedding space. For each topic $t$, every word $w^t$ within that topic was assigned a probability ($p_w^t$) by BERTopic, indicating its contribution to the overall representation of the topic. To compute an embedding for a topic ($\overrightarrow{t}$), we calculated the weighted sum of the embeddings of its topic words, where each word's embedding was weighted by its normalized probability ($p_w^t$):

\begin{equation} \label{topic-emb-equation}
\overrightarrow{t} = \sum_{w^t} \text{emb}(w^{t}) \cdot p_w^t, 
\end{equation} where \emph{emb} is the sentence transformer model that maps the topic words into embeddings, and $w^t$ are all the words associated with topic $t$.

To facilitate semantic clustering of the topics, we standardized their embeddings and reduced their dimensionality from 768 to 10 using UMAP. We then explored two clustering algorithms, K-Means and DBSCAN, applied to the reduced topic embeddings. Through manual inspection, we observed that the K-Means clustering method effectively generated semantically coherent clusters, with most topics within each cluster sharing a cohesive theme. We also observed that K-Means with 20 clusters produced the result with the best quality. Further details on the analysis can be found in Section SI2.1.

\subsubsection*{Filtering non-dream content and adjusting themes}\label{sec:topic-classification}

K-Means give us 20 clusters that contained dreaming and non-dreaming topic clusters, as well as a mixture of both. Despite the potential inclusion of non-relevant elements in dream reports, BERTopic's remarkable semantic capabilities proved highly effective in distinguishing the actual recollections from other content. Leveraging this capability, we employed a filtering process to exclude non-dream themes and topics from our analysis.

To achieve this, we manually classified each theme into one of five categories:
\begin{enumerate}[itemsep=0pt]
\item Dream-related content -- encompassing elements such as dream locations, animals, and people.
\item Dream types -- including nightmares and recurring dreams.
\item Dreaming-waking interface topics -- covering experiences like waking up crying or feeling confused, as well as being awakened by an alarm.
\item Waking phenomena -- addressing aspects such as mental health issues and the date and time of the dream.
\item Social media artifacts -- encompassing elements like expressions of gratitude to the reader or requests for dream interpretation.
\end{enumerate}
To assign these categories, we relied on the 10 topic words associated with each topic within a theme. In cases where necessary, we also consulted three representative dream sentences generated by BERTopic, along with 20 randomly sampled dream sentences. In most instances, the topics naturally fell into one of the predefined categories, allowing us to filter out entire themes comprising non-dream content. 

We found three non-dream clusters: \emph{Dream types}, \emph{Dreaming-waking interface topics}, and \emph{Social media artifacts} as discussed in the above. Notably, \emph{Waking phenomena} was not present as a coherent cluster. We had to manually inspect all the 288 topics and top corresponding dream sentences, which allowed us to separate out this category. 
For example, from the theme \emph{Time, Time Travel, and Timelines}, we excluded topics such as \emph{`5am currently clock checked phone'} and \emph{`date 2018 June.'} Similarly, in the theme \emph{Mental Reflection and Interactions,} we omitted topics such as \emph{`interpret, does really mean'} and \emph{`don remember details.'}
After this filtering step, we ended up with the final 217 dream topics.

In addition, some composite clusters contained divergent dream themes (i.e., semantically dissimilar overarching themes). For instance, there was a cluster talking about \emph{Outdoor environment and space}, which could be deconstructed into three themes: \emph{Outdoor locations}, and \emph{Space} and \emph{Weather}. So, we manually split up such clusters (and also carried out minimal re-adjustment of topics into appropriate clusters (e.g., our largest topic (\emph{0  lady, face, looked, head}) was incorrectly present in a cluster which we later termed \emph{Other topics}, and which is as a catch-all for topics which did not fit into a semantically coherent cluster. We moved this topic into \emph{People and relationships}). 

After carrying out this whole process, we end up with 217 topics clustered into 22 dream themes.

\subsection*{Building dream topics taxonomy (co-occurrence network)}\label{sec:topic-cluster-freq}

To create a taxonomy of dream topics, we employed a network-based approach that explores the interplay between dream themes and their constituent topics within dream reports. To facilitate this analysis, we conducted an initial assessment of the frequency distributions of both topics and themes across the entire corpus of dream reports.

For topics, we linked each one to a dream report (/dreamer) if the topic was found at least once in the report (in all dream reports of that dreamer), and counted the number of dreams (/dreamers) associated with each topic. In Table~\ref{tab:topics}, we present the topic name, top 4 topic words, number of dreams, and number of dreamers for the top 20 most frequent topics. Full statistics (e.g., the 10 topic words and number of sentences associated with each topic) for these and the rest of the topics, can be found in the Supplementary Information File (see Data Availability section). 

Similarly, we linked a dream report (/dreamer) to a theme, if any of the theme's constituent topics was found at least once in the report (/in all dream reports of that dreamer). Additionally, we computed the number of topics in each theme associated with corresponding dream reports (dreamers). In Table \ref{tab:topic_clusters}, we present the theme name, top 3 topics associated with it, the total number of topics in it, and number of dreams, and number of dreamers associated with it. Additionally, we linked to a corresponding \emph{HVdC} category each theme for which such a link is found. The full list of topics belonging to each theme can be found in Table SI1.

Having these frequencies at hand, we built the co-occurrence network of dream themes as follows. Each node in the network represents a theme, and pairs of themes were connected by an edge if they co-occurred in a dream report. The edges were weighted by the number of dream reports in which such a co-occurrence was found. This undirected network had a single connected component with 22 nodes and edge weights ranging from 13 to 7643 (Mean = 762.31 $\pm$ 928.54 and median = 482.0).

For the purpose of visualization, we used backboning to sparsify the network by preserving the most important edges. We used noise-corrected backboning~\cite{coscia2017network}---a technique that relies on a statistical null-model to identify and prune non-salient edges---with a backboning threshold of 3.8 (which reduced the network from $231$ to $46$ edges). We used Gephi~\cite{bastian2009gephi} to visualize this network (see Figure \ref{fig:taxonomy}). We scaled the size of the nodes according to the number of dreams associated with each theme.

\subsection*{Finding topics and themes by dream type (odds-ratio analysis)}
\label{sec:odds-ratios}
In addition to discovering common topics across all dreams, we studied whether specific types of dreams (i.e., nightmares, lucid, vivid, and recurring dreams) are characterised by particular topics and themes. 
Odds-ratio metric allowed us to do so as it compares the odds of a topic occurring in the specific type of dreams (e.g., recurring) to the odds of the same topic occurring in the rest of dreams. We first assigned the dream reports to the corresponding dream types by searching for relevant keywords (‘nightmar’, ‘recurring’ or ‘re-occurring’,  ‘lucid’, ‘vivid’) in Reddit post title and body or, if the dream type had a corresponding flair (which were present for nightmares and recurring dreams only). If either of these conditions were satisfied, we assigned the dream to the experimental subset; else to the control subset. 

First, we defined a topic or theme $t$'s importance in a dream $d$ as:
\begin{equation}
{ 
I_{t,d} = \frac{\# \text{ sentences mentioning topic t in d}}{\# \text{ total sentences in d}} 
} \label{eq:topic_importance}
\end{equation}

We then computed the odds ratio for all topics and themes across the 4 dream types as follows:

\begin{equation}
\text{Odds Ratio (DT, t)} = \frac{\text{Odds of DT association with t}}{\text{Odds of the rest of dreams association with t}} =
\frac
{
\frac
{ \sum_{d \in DT} I_{t,d}}
{\text{\# dreams in DT that do not contain t}}
}
{
\frac
{ 
\sum_{d \notin DT} I_{t,d}}{\text{\# dreams that are not in DT that do not contain t}}
}
\end{equation}
where DT is a dream type, t is a topic or theme.

\subsection*{Finding topics and themes through time}\label{sec:temporal-analysis}
Analyzing the temporal dynamics of dream topics and themes holds particular significance, especially in light of major events such as the COVID-19 pandemic, known to have a substantial impact on collective dreaming patterns.~\cite{barrett2020dreams,vscepanovic2022epidemic,schredl2020dreaming}

For our analysis, we focused on a monthly timescale, leveraging data spanning from January 2019 to August 2022, encompassing a total of 44 months. The inclusion criterion for each month required a minimum of 300 dream reports, ensuring robust statistical representation. Throughout this period, February 2019 recorded the lowest number of dreams ($n = 366$), whereas January 2021 exhibited the highest dream count ($n = 1573$). The average number of dreams per month was $925 \pm 332$, with a median of $935$ dreams per month.

We used topic/ theme importance in a dream ($I_{t,d}$) introduced in Equation \ref{eq:topic_importance} to calculate topic/ theme importance at a given point in time $m$ (i.e., month) as follows:
\begin{equation}
    I_{t,m} = \frac{\sum_{(\text{d posted at time $m$})} I_{t,d} }{\text{\# dreams posted at time $m$}}
\label{eq:topic_importance_avg}
\end{equation}

We tracked z-scores of topic/ theme importances $I_{t,m}$ through time, to understand the relative change of a topic/ theme w.r.t. itself:
\begin{equation}
    z_{I_{t,m}} = \frac{I_{t,m} - \mu_{I_{t,m}}}{\sigma_{I_{t,m}}}
\label{eq:zscore}
\end{equation}
To additionally improve the quality of signals, we used the centered average with a window length of 5 for smoothing the temporal plots:
\begin{equation}
    \widetilde{z}_{I_{t,m}} = \sum_{t=(t-2)}^{(t+2)}{z_{I_{t,m}}}.
\label{eq:zscore_smooth}
\end{equation}

\section*{Acknowledgements}
L.M.A acknowledges the support from the Carlsberg Foundation through the COCOONS project (CF21-0432). The funder had no role in study design, data collection and analysis, decision to publish, or preparation of the manuscript.

\section*{Author contributions statement}

 A.D., S.\v{S}, and L.M.A. conceived and designed the experiment(s),  A.D. conducted the experiment(s), A.D., S.\v{S}, L.M.A., R.M., D.B., and D.Q. analysed the results. All authors wrote and reviewed the manuscript. 

\section*{Data Availability Statement}\label{sec:data-availability}
The data generated in this study is publicly available and can be accessed at \href{https://doi.org/10.6084/m9.figshare.23618064.v2}{https://doi.org/10.6084/m9.figshare.23618064.v2}.

\newpage

\section*{Supplementary Information}

\renewcommand{\thefigure}{SI\arabic{figure}}
\setcounter{figure}{0}
\renewcommand{\thetable}{SI\arabic{table}}
\setcounter{table}{0}

\section{Data statistics (Extended)}
\begin{figure}[h]
\centering
\includegraphics[width=0.5\textwidth]{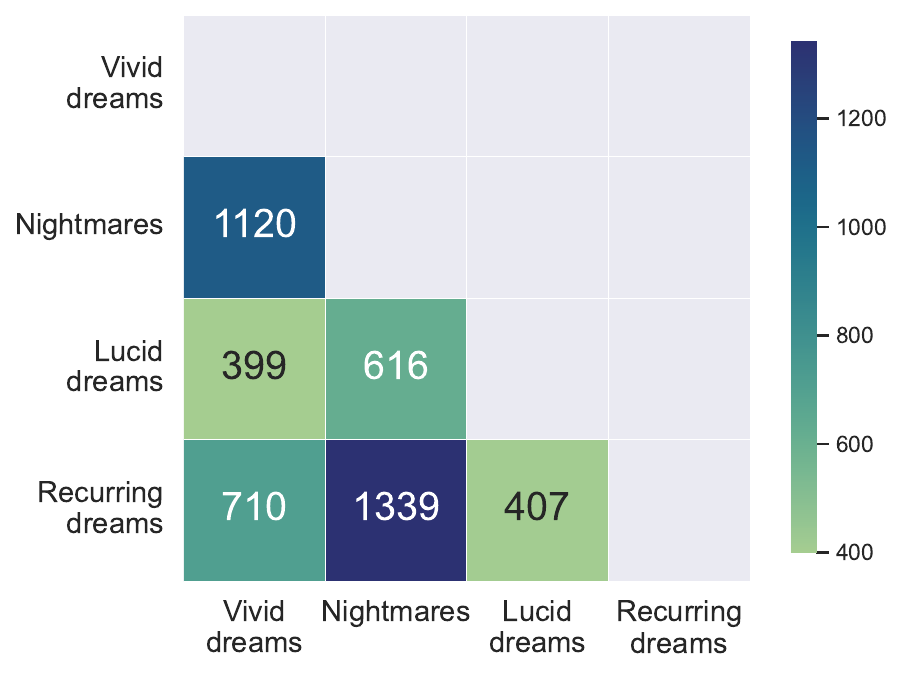}
\caption{\textbf{Heatmap that shows no. of dreams present across two dream types simultaneously} \label{fig:num_dreams_heatmap}}
\end{figure}

\begin{figure}[h]
\centering
\includegraphics[width=0.9\textwidth]{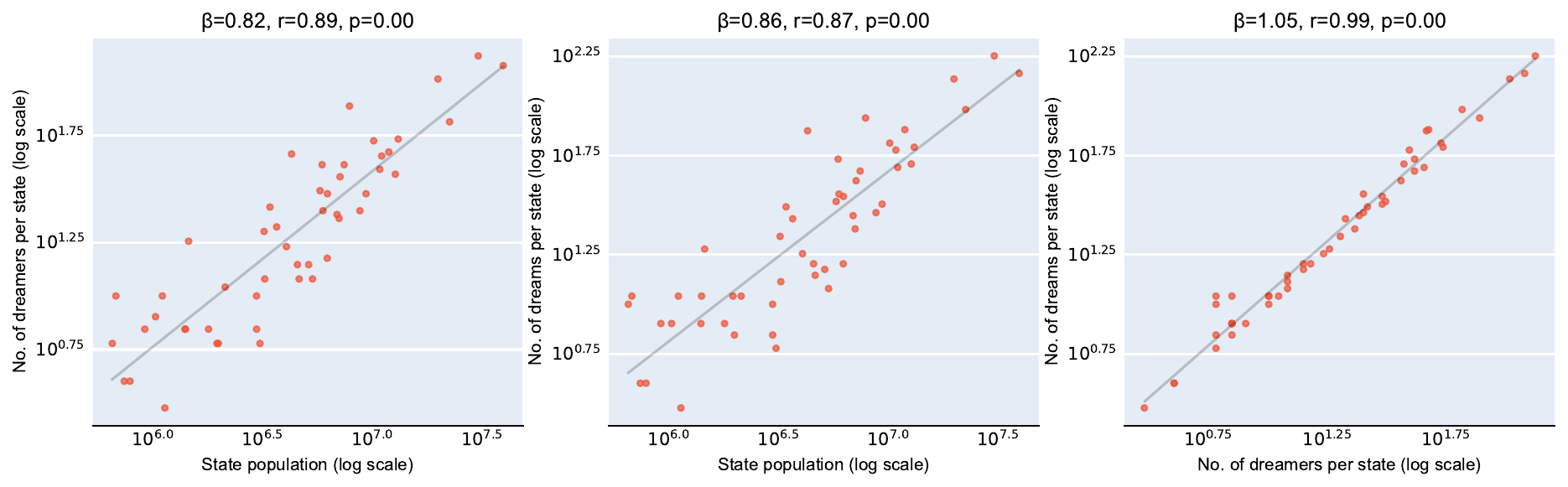}
\caption{\textbf{Correlation plots b/w 1. US state population, 2. No. of dreamers per state, and 3. No. of dreams per state} \label{fig:us_correlation}}
\end{figure}

\begin{figure}[h]
\centering
\includegraphics[width=0.9\textwidth]{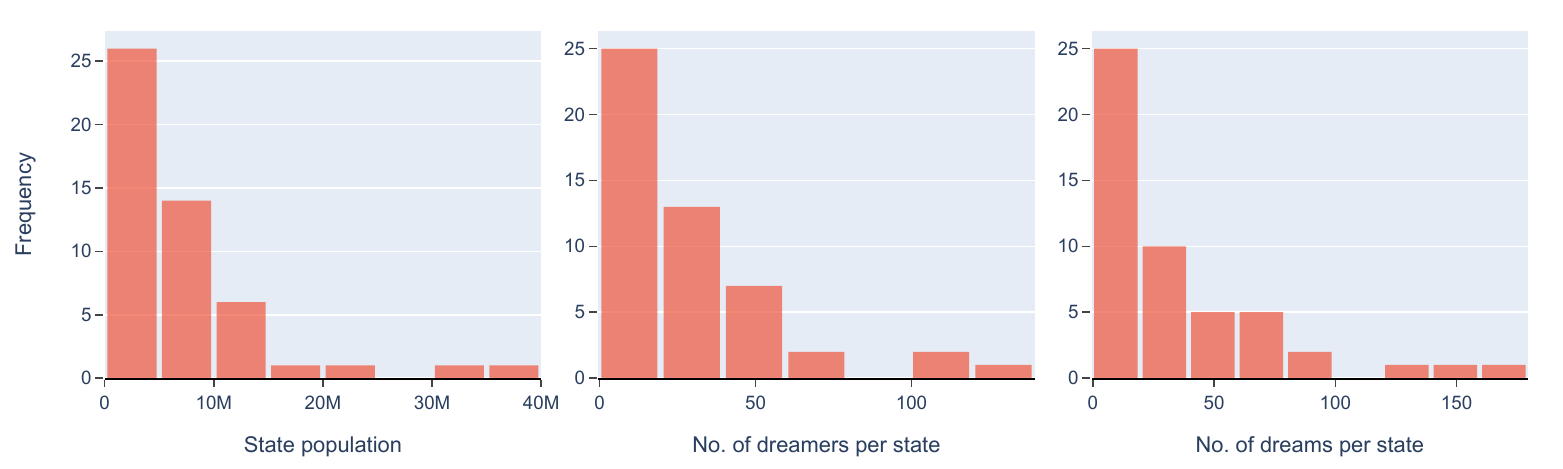}
\caption{\textbf{Distribution of state population \& no. of dreams, dreamers per state} \label{fig:users_state_distrib}}
\end{figure}

\begin{figure}[h]
\centering
\includegraphics[width=0.9\textwidth]{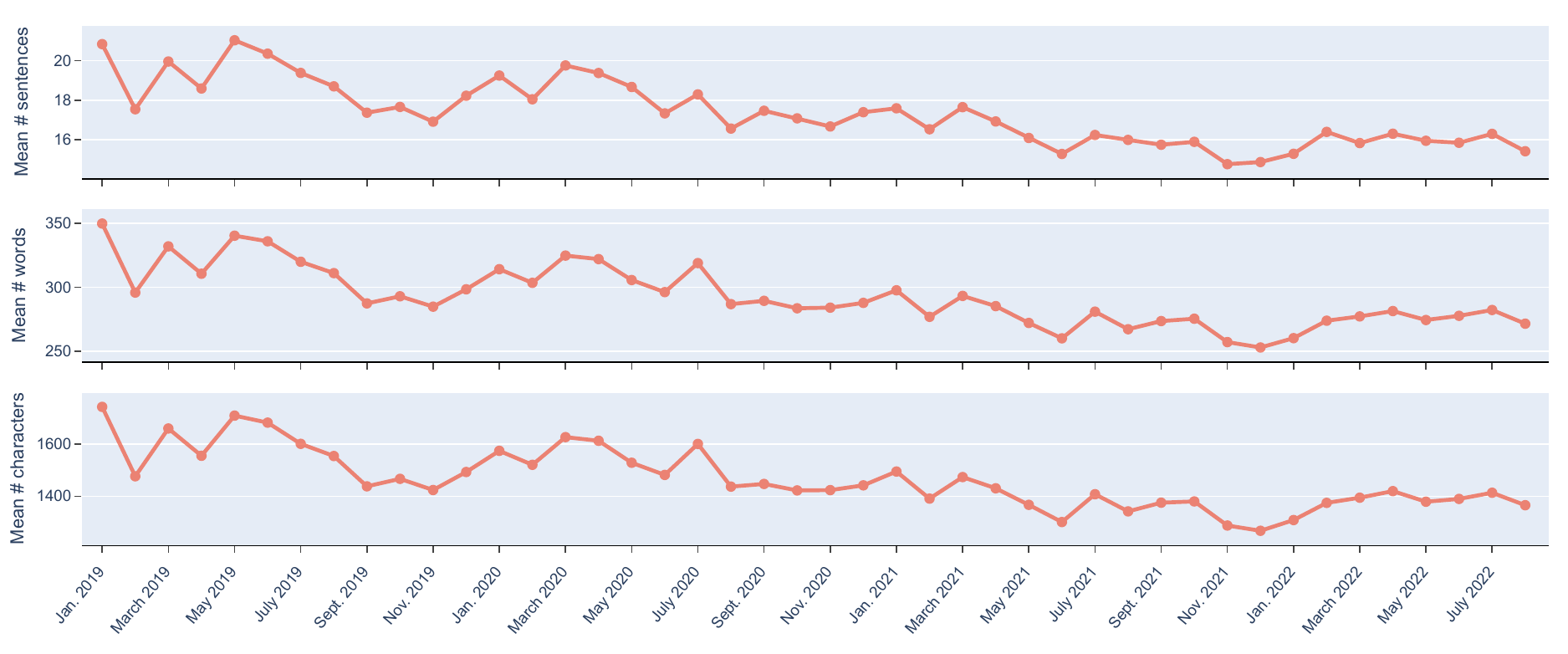}
\caption{\textbf{Evolution of dream length over time} \label{fig:dream_len_temporal}}
\end{figure}

Fig. \ref{fig:num_dreams_heatmap}, \ref{fig:us_correlation}, \ref{fig:users_state_distrib} \& \ref{fig:dream_len_temporal} provide additional insights into the dream reports. We were able to geolocate 4.04\% (n=1784) of all dream reports (distributed amongst 3.22\% (n=1423) users) at the US state level. Fig. \ref{fig:us_correlation} demonstrates representative coverage of our data in all 50 US states; using the aforementioned data subset.

\section{Methods \& Implementation Details (Extended)}

\subsection{Details of Topic Modelling}

BERTopic embeds all the documents into vectors, projects them onto a lower dimensional space using UMAP, clusters the reduced embeddings using HDBSCAN (Hierarchical Density-based Spatial Clustering of Applications with Noise), and finally generates the topic names using a class-based variation of TF-IDF, namely c-TF-IDF.

In the application of BERTopic, we set the hyperparameters of the \emph{all-mpnet-base-v2} sentence transformer model such that the min. no. of dream sentences required to form a topic is 100 (min\_topic\_size = 100) and it  automatically merges similar topics together using HDBSCAN (Hierarchical Density-Based Spatial Clustering of Applications with Noise) (nr\_topic = `auto'). 

After fitting the model, we used its functionality to update the topic representation i.e., modify the topic words to remove English stop words and take into account both unigrams and bigrams as topic words/phrases. This is particularly important for 2-word phrases like \emph{sleep paralysis}, \emph{recurring dreams} and, \emph{serial killers} which we found, quite frequently mentioned in the extracted dataset of topics.

The probabilities of all topic words for each topic sum to 1; however since we removed ``dream'' and ``dreams'' topic words, we had to re-normalize the topic word probabilities to ensure they summed to 1. 

\begin{figure}[!h]
\centering
\includegraphics[width=0.9\textwidth]{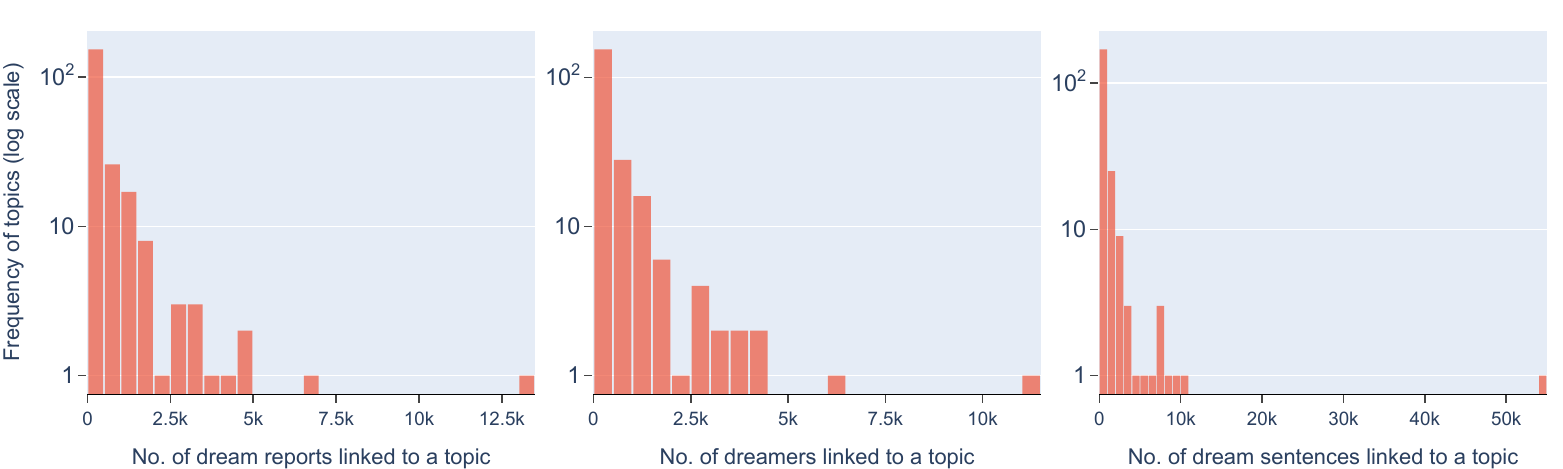}
\caption{\textbf{Distribution of no. of dreams, dreamers \& dream sentences linked to a topic} \label{fig:topics_distrib}}
\end{figure}

Fig. \ref{fig:topics_distrib} captures the distribution of no. of dreams reports, dreamers \& the no. of dream sentences linked to a topic. There were 1982 (4.48\%) dreams distributed amongst 1277 dreamers which weren't associated with any topic at all. While there were 4978 (11.26\%) dreams amongst 3416 dreamers which weren't associated with any dreaming topics or themes.

\begin{figure}[!h]
\centering
\includegraphics[width=0.98\textwidth]{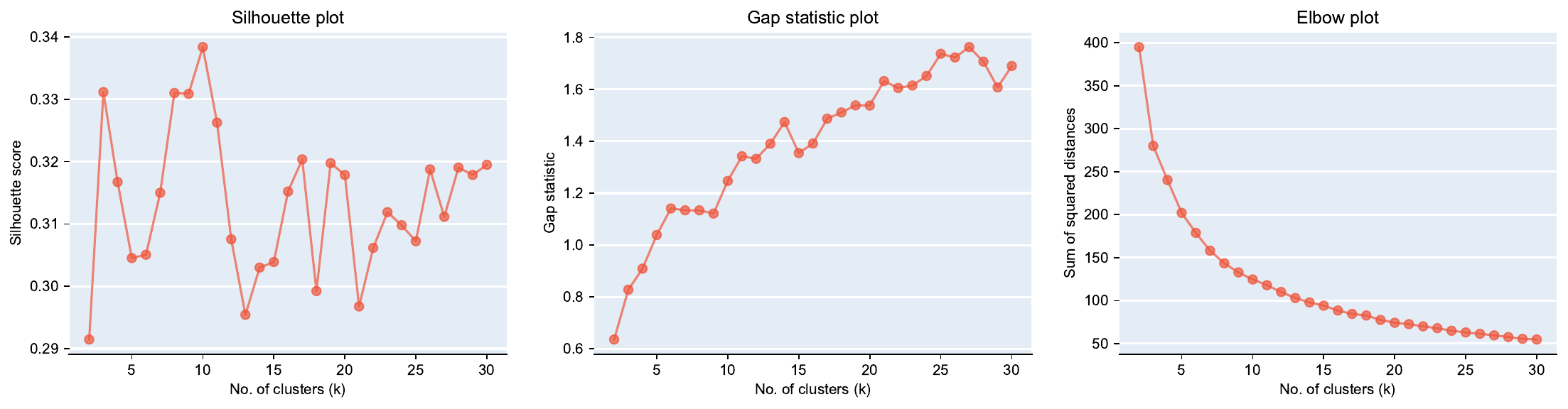}
\caption{\textbf{Selection of no. of clusters (k) for K-Means} \label{fig:k_means_clusters}}
\end{figure}

Fig. \ref{fig:k_means_clusters} shows the plots for selecting the no. of clusters, post application of K-Means to topic embeddings.

\begin{figure}[!ht]
\centering
\includegraphics[width=0.5\textwidth]{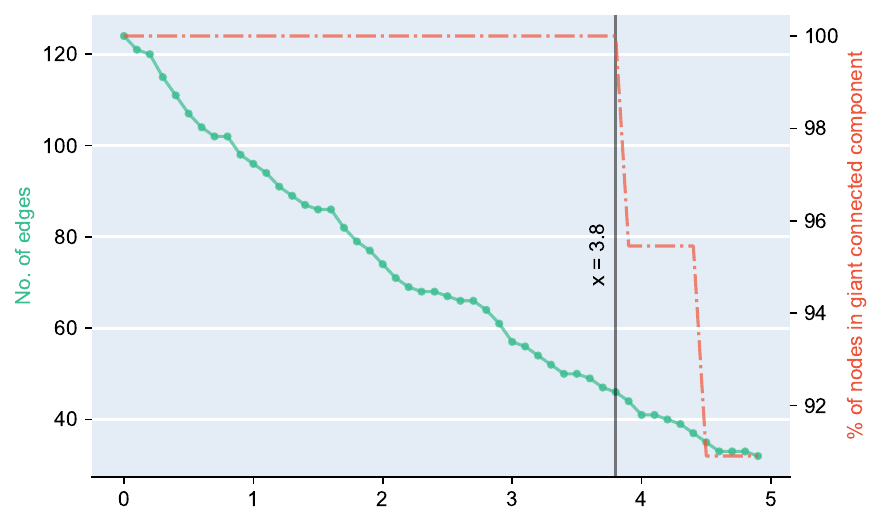}
\caption{\textbf{Backboning graph for theme network with backboning threshold = 3.8} \label{fig:backboning_plot_theme_net}}
\end{figure}

\subsection{Details of building dream topics taxonomy (co-occurrence network)}
Fig. \ref{fig:backboning_plot_theme_net} shows the plot used to determine the backboning threshold for visualizing the theme co-occurrence network.

\subsection{Details of finding topics and themes through time}

We did not include dreams from September 2022 (last month in our dataset) in temporal analyses as we only collected data up until 7th September 2022. Manual inspection revealed that, during 2018, the temporal curves for the raw z-scores were quite noisy to accurately infer anything, hence we further restricted our analyses from January 2019 to August 2022. For July and August 2022, since we did not have the data from the subsequent months for smoothing, we used $min\_periods = 0$ for the $rolling$ function in Pandas which helped to compute the average by excluding the subsequent months' data. For calculating the smoothed z-scores of January and February 2019, we leveraged the values from November and December 2018, prior to discarding them, as discussed above.

\section{Results (Extended)}
Table~\ref{tab:supp-all-topics-themes} provides an exhaustive list of dream topics and themes they belong to; as discovered by our proposed methodology.

\begin{table}[]
\tiny
\centering
\caption{Full list of dream topics binned into various themes\label{tab:supp-all-topics-themes}}
\begin{tabular}{p{.02\linewidth}p{.05\linewidth}p{.8\linewidth}}
\toprule
\multicolumn{1}{l}{\textbf{Theme no.}} & \textbf{Theme} & \textbf{Topics} \\ \midrule
0 & People and relationships & 0 lady, face, looked, head; 2 dream, girl-dream, dreamt, ve; 8 ex, years, dating, talk; 11 dreams-mean, staring, talk, people-dream; 16 dream-dad, dad-dream, dream-mom, mom-dream; 19 mum, aunt, mom-came, called-mom; 80 ex-boyfriend, dreams, having-dreams, relationship; 92 hugged, hug, kissed, arms; 106 names, remember-faces, characters, man-woman; 107 ll, nickname, named, mr; 115 crush, falling-love, fell-love, fall-love; 117 sex-dream, sexual, wet-dream, dream-sex; 169 versions, different-person, identity, perfect; 186 relationship, brother, mother-father, wife; 213 cheating, having-dreams, dream-boyfriend, relationship; 243 brother-dream, younger-brother, dream-younger, talking-little; 258 clowns, little-girl, creepy, costume \\

\arrayrulecolor{black!20}\midrule1 & Indoor locations & 5 mall, restaurant, eating, ice-cream; 7 bus, driving, cars, train-station; 13 doors, house, rooms, mansion; 36 toilet, shower, restroom, bathrooms; 37 house-dream, dream-house, apartment, childhood-home; 42 elevator, stairs, floors, escalator; 50 hospital, surgery, nurse, doctors; 60 flying, airport, helicopter, planes; 116 universes, alternate-reality, different-dimension, travel; 147 911, dial, emergency, ambulance; 170 portals, dimension, tunnel, hell; 220 roller-coaster, rollercoaster, rides, wheel; 250 house, amp-x200b, window, walking; 273 maze, hyper, tunnels, runner; 275 tier, different-levels, sam, happens-time; 277 elevators, dreams, reoccurring-dreams, recurrent \\

\arrayrulecolor{black!20}\midrule2 & Violence and death & 14 death, died-dream, going-die, death-dream; 43 pistol, shooting, shot-head, shotgun; 44 police, officer, officers, kidnapped; 46 nuke, war, fires, missile; 49 soldiers, hitler, russia, world-war; 76 serial-killer, dream-killing, killed-dream, killers; 83 tw, warning, nsfw, sexual-assault; 101 passed-away, cancer, away-years, father-passed; 156 rape, sexually-assaulted, abuse, nightmares; 158 jail, dream, court, police-car; 205 fight, vs, duel, eachother; 209 violent, disturbing, dreams-past, weird-dreams; 238 killed, putting-hands, fighting, beat; 268 violence, person-real, want-hurt, hate; 279 intruder, breaking, recurring, recurring-dream \\

\arrayrulecolor{black!20}\midrule3 & Mental reflection and interactions & 26 know-make, sure-don, know-think, know-sure; 40 yeah, like-wtf, hell, like-fuck; 52 asked-doing, tell, help, begged; 53 know, okay, yeah, know-going; 61 don-want, needed, doing-don, remember-doing; 120 said-yes, respond, didn-answer, yes-did; 122 know-knew, knew-didn, knew-don, known; 127 matter, case, know-isn, mattered; 139 lied, did-don, wish, don-just; 151 plan, care, notices, did-just; 161 confused-just, im, kinda-confused, just-really; 175 couldn-just, couldn-time, couldn-didn, know-able; 195 make-sense, makes, logic, sense-like; 217 miss, say-love, loved, dearly; 218 remember-saying, said-words, understand, mumbling; 233 oh, say, reply, asked-said; 235 doesn-exist, didn-exist, just-know, indication; 253 realised, exploring, felt-confused, bathroom-looked \\

\arrayrulecolor{black!20}\midrule4 & Feelings & 20 felt-real, dream-felt, real-dream, real-like; 39 woke-crying, started-crying, woke-tears, crying-dream; 64 feel-pain, painful, pain-dream, felt-pain; 75 relieved, felt-peace, happiness, felt-happy; 81 expression, tears, shock, nervous; 119 wasn-scary, scary-just, frightening, scariest; 129 ve-felt, felt-feeling, feeling-felt, haven-felt; 132 feel-right, felt-wrong, wasn-right, feel; 134 shocked, shock, surprise, disbelief; 136 crying, gasped, terrified, fear; 160 felt-sad, kind-sad, feel-sad, distraught; 172 upset, angry, livid, got-pissed; 194 guilt, felt-guilty, feel-guilty, remorse; 201 panic, panicking, panicked, started-panic; 242 really-creepy, weird-creepy, creepiest, sounds; 266 calm, calmed, tried-calm, feel-calm \\

\arrayrulecolor{black!20}\midrule5 & Sights and vision & 4 lights, sun, pitch-black, wearing; 87 reflection, looked-mirror, looking-mirror, mirrors; 140 blinded, vision-blurry, blur, recite; 174 disappeared, went-look, window, lost; 211 faded-away, existence, outline, suddenly-felt; 212 stared, head-looked, looked-eyes, looking-direction; 215 blurry, quality, picture, coated; 222 sight, seen, walkway, bodies; 226 sight, majestic, really-pretty, stunning; 259 mirrors, looking-mirror, bathroom-mirror, dream-looked; 264 decided-look, kept-looking, look, looking-looked \\

\arrayrulecolor{black!20}\midrule6 & Animals & 10 kitten, lion, birds, owl; 48 spider, maggots, batman, thanos; 88 snakes, alligator, turtle, bite; 176 horse, goats, brown, granny; 191 bear, polar, fence, started-chasing; 203 wolf, pack, grey, attack; 241 bees, stung, swarm, buzzing; 263 rat, traps, bites, swarming; 265 monkeys, enclosure, jungle, goo; 267 dinosaur, park, spawn, edges; 276 bunny, glitched, grey, hopping; 282 toad, poison, green, psychedelic; 284 deer, saw, tried-bite, road \\

\arrayrulecolor{black!20}\midrule7 & Other topics (size, smell, apocalypse, etc.) & 25 dream-ends, story, endings, recurring-theme; 86 huge, inches, like-size, big-small; 99 pov, 3rd-person, person-perspective, person-view; 111 room-starts, walking, family, house-starts; 112 really-cool, pretty, neat, thought-cool; 118 apocalyptic, dream-end, world-dreams, earth; 142 look-like, hair, skinny, face; 153 thousands, total, maybe-20, voodoo; 166 apocalypse, world-end, cosmic, event; 173 gay, trans, gender, feminine; 181 started-normal, normal-like, normal-looked, completely-normal; 188 smell, smelled-like, rot, disgusting; 199 describing, hard, try-best, properly; 206 coincidence, connection, thought, uncanny; 207 scene-changed, things-changed, switched, shifting; 214 chaos, disaster, mess, society; 216 dream-like, like-ve, haven-dream, like-long; 227 horrible, good-bad, bad-just, know-good; 239 knew, travelling, end-finally, didn-finish; 261 know-sounds, sounds-weird, weird-know, crazy; 262 scientists, experiments, laboratory, discover \\

\arrayrulecolor{black!20}\midrule
8 & Outdoor locations & 12 beach, ocean, swim, river; 29 path, garden, hills, plants; 105 cave, tunnels, underground, tower; 114 forest, field, grass, dream-starts; 202 wooden, driving, bridges, unfinished; 249 islands, beings, coast, plans \\

\arrayrulecolor{black!20}\midrule
9 & Personal objects & 33 dressed, mask, naked, clothing; 56 phones, ringing, check-phone, battery; 74 pages, pen, ink, letters; 143 boxes, necklace, rings, cardboard-box; 145 shoes, pair, barefoot, sock; 146 crystals, statues, stone, gems; 163 backpack, bags, suitcase, packing; 164 knives, axe, pocket-knife, kitchen; 177 projector, remote, screen, television; 180 dolls, porcelain, voodoo, haunted; 193 counting, lottery, significance, noticed; 224 blankets, bed, pillow, felt; 246 paintings, art, wall, canvas; 255 eggs, ipad, scrambled, cracking; 270 keys, pocket, puzzle, box; 274 tattoos, artist, sleeve, piercing; 281 coins, drawer, wallet, rare \\
\arrayrulecolor{black!20}\midrule
10 & School & 15 classroom, teachers, principal, student; 27 school-dream, dream-school, college, dream-high; 269 tests, failed, exam, professor \\
\arrayrulecolor{black!20}\midrule
11 & Movement and action & 67 left, wanted-leave, time-leave, wanted-home; 85 continued-walking, continue, street, walk-home; 95 run, started-running, run-like, sprinting; 98 falling, hit-ground, fall-ground, let-fall; 124 lungs, couldn-breathe, like-couldn, heart; 128 remove, string, try-pull, tugged; 135 control, couldn-control, control-body, hopeless; 141 hid, closet, stairs, underneath; 144 escape, escaped, way-escape, managed-escape; 237 lose, round, hockey, scissors; 272 queue, line-people, waiting-line, standing-line; 280 dream-running, running-dream, run-fast, dream-run \\
\arrayrulecolor{black!20}\midrule
12 & Supernatural entities & 28 creature, bite, mouth, face; 96 shadows, shadowy-figure, dark-figure, entity; 97 zombies, zombie-apocalypse, outbreak, like-zombie; 125 aliens, invasion, grey, race; 178 dragon, chinese, bearded, red; 184 alien, invasion, abducted, jar; 189 robots, ai, giant, colony; 192 zombie-apocalypse, zombies, outbreak, walking-dead; 197 ghosts, haunted, like-ghost, saw; 278 vampires, gang, hunter, glowing-red; 283 clones, disabled, machine, scientist \\
\arrayrulecolor{black!20}\midrule
13 & Sounds and lack thereof & 41 singing, songs, lyrics, stage; 77 voices, heard-voice, hear-voice, voice-head; 89 footsteps, ghost, noises, ringing; 100 paralyzed, able, speak, couldn-speak; 121 silent, sound, went-silent, eerily; 182 speak, couldn-speak, mouth, whisper; 223 laughter, couldn-hear, hear-talking, couldn-understand \\
\arrayrulecolor{black!20}\midrule
14 & Media and tech & 31 theater, anime, movie-like, tv; 57 minecraft, games, vr, game-like; 90 game-dream, dream-playing, vr, video-games; 219 cartoon, documentary, girl-guy, lisa; 236 batman, marvel, thanos, loki; 285 spongebob, episode, freeze, robots \\
\arrayrulecolor{black!20}\midrule
15 & Religious and spiritual & 66 demons, devil, monster, demonic; 68 church, cult, lot-people, dream-god; 70 demon, devil, demons, angel; 109 religious, atheist, catholic, supernatural; 110 hell, purgatory, afterlife, heaven-hell; 137 powers, abilities, superpowers, telekinesis; 229 cult, leaders, satanic, ritual \\
\arrayrulecolor{black!20}\midrule
16 & Life events & 59 birth, babies, pregnancy, newborn; 72 party, invited, having-party, brother; 78 giving-birth, dreamt, dream, twins; 190 party, party-dream, dream-party, dinner; 198 wedding, ceremony, aisle, venue; 252 wedding, getting-married, engaged, ceremony \\
\arrayrulecolor{black!20}\midrule
17 & Human body, especially teeth and blood & 35 blood, skin, humanoid, arms; 108 teeth, falling, tongue, pain; 130 teeth, tooth, falling, gums; 133 blood, like-blood, covered-blood, splatter; 254 bodies, corpses, dead-bodies, body-parts \\
\arrayrulecolor{black!20}\midrule
18 & Work & 47 new-job, boss, jobs, shift \\
\arrayrulecolor{black!20}\midrule
19 & Weather, especially storms & 79 rain, tornado, hurricane, started-raining; 104 snow, snowing, cold, winter; 271 tornado, storms, april, category \\
\arrayrulecolor{black!20}\midrule
20 & Time, time travel and timelines & 102 timeline, time-travel, time-skip, like-time; 165 time-travel, dream-world, universes, dream-time; 187 noon, 00am, early-morning, evening \\
\arrayrulecolor{black!20}\midrule
21 & Space & 159 sun, earth, eclipse, phases; 221 space, space-ship, nasa, oxygen; 251 meteors, earth, asteroid, coming \\ 
\arrayrulecolor{black}\midrule
\end{tabular}
 \end{table}

\end{document}